\documentclass[11pt,a4paper]{article}
\pdfoutput=1  
\usepackage{epsfig}
\usepackage{graphicx,psfrag}
\usepackage{multirow}
\usepackage{slashed}
\usepackage{pstricks}
\usepackage{caption}
\usepackage{subcaption}
\usepackage{url}
\usepackage{braket}
\usepackage{jheppub}
\usepackage{enumerate}
\usepackage{wasysym} 
\usepackage{mathrsfs} 
\usepackage{amsfonts} 
\usepackage{amsbsy} 
\usepackage{amscd}
\usepackage{color}
\usepackage{changepage}
\usepackage{floatrow}
\usepackage{makecell}
\usepackage[utf8x]{inputenc}

\makeatletter

\def\eeq{\end{equation}}
\def\beq{\begin{equation}}

\newcommand{\Rmnum}[1]{\expandafter\@slowromancap\romannumeral #1@}

\newcommand{\bea} {\begin{eqnarray}}
\newcommand{\eea} {\end{eqnarray}}

\newcommand{\gsim}{\raisebox{-0.13cm}{~\shortstack{$>$ \\[-0.07cm]
      $\sim$}}~}

\makeatother

\title{Analyzing polarized boosted top from stop decay at high luminosity and high energy LHC}

\author[a]{Jayita Lahiri}
   \affiliation[a]{Department of Physics, Indian Institute of Technology Guwahati,
North Guwahati, Assam - 781039, India}

\emailAdd{jayitalahiri@rnd.iitg.ac.in}

\abstract{
Measuring top polarization is extremely important in understanding any new physics that couples to top quark. Here, we take up the task of extracting polarization of top quark coming from TeV-scale stop decay at the upgraded high luminosity and high energy LHC. We focus on the scenario where the two tops from the stop decay lead to semi-leptonic final state. We take into account all relevant backgrounds and put kinematical cuts to suppress them. We arrive at a signal region where the desired signal can be observed with reasonable significance. Next, we examine, in such signal regions, how well top polarization can be measured. For that we have calculated the degree with which left- and right-handed top quarks can be separated from each other, in terms of $P$-values. We have also performed a comparative study between various centre-of-mass energies and integrated luminosities in terms of their reach in probing top polarization from stop decay.

}

\begin{document}

\maketitle

\section{Introduction}

Top quark physics is one of the most promising sectors of the Standard Model(SM). Due to its large mass, on one hand, it can have close relation with the electro-weak symmetry breaking. On the other hand, it works as an important probe of new physics(NP) at the TeV scale. Because of its large mass it also has a small enough life-time, allowing it to decay before hadronization. Studying its decay products gives us direct access to the properties of top quark. For example, in the context of SM, top quark coupling with Higgs boson, its magnitude as well as Lorentz structure can be examined looking at the decay products of top~\cite{CMS:2020djy}. Top coupling with Higgs has been used to measure the CP-properties of the Higgs boson~\cite{BhupalDev:2007ftb,Ananthanarayan:2013cia,Ananthanarayan:2014eea}. On the other hand, most TeV-scale theories offer a top partner, whose coupling with the top quark in general can follow interesting chiral structure. There can also be heavy gauge bosons, which can give rise to polarized top quarks. Looking at top quark polarization thus gives a handle to probe the coupling structure and allows us to get a deeper understanding about the high-scale physics. At the same time, exploring such couplings can be used to constrain the NP. There is a plethora of such models which gives rise to a polarized top. One can recall $W^\prime$ or $Z^\prime$ models, supersymmetric models such as Minimal Supersymmetric Standard Model(MSSM), extra-dimensional models with Kaluza-Klein states to name a few.

Polarization of top quark depends on the top quark interactions at the production level. For example, in SM $t \bar t$ pair is produced via parity conserving QCD interaction and therefore the top quark produced is mostly unpolarized in that case. On the other hand, single top quark is produced mainly via weak $V-A$ interaction, and therefore, is primarily left-handed. However, one should also note that, even in case of unpolarized top, such as QCD production of $t \bar t$ pair, the spin correlation between the top-antitop pair is non-zero. Therefore the spin-correlation in principle plays an important role alongside top-quark polarization, in the context of probing the coupling structure. The spin correlation has been studied in detail in~\cite{Barger:1988jj,Uwer:2004vp,Mahlon:1995zn,Bernreuther:2001rq,Baumgart:2012ay,CDF:2010yag,CMS:2016piu,ATLAS:2014abv}. 
On the other hand, single top production in SM will give rise to left-handed tops, because of its production via parity-violating $V-A$ theory. Any deviation from this SM chiral structure would in principle be a hint for new physics.

Therefore, the measurement of top polarization becomes imperative in the studies of polarization-sensitive NP models. Top quark polarization has been measured in various distributions of the charged lepton from the top quark decay~\cite{Jezabek:1994qs,Grzadkowski:2001tq,Godbole:2006tq,Godbole:2007zzb,Godbole:2010kr,Hagiwara:2017ban,Jueid:2018wnj}. 
However, when top is produced via the decay of some heavy state, it will be boosted and its decay products will be collimated in the form of a fatjet. In such scenarios, leptonic final state becomes less effective in measuring top polarization.
The measurement of top quark polarization in the hadronic final state has been explored~\cite{Kaplan:2008ie,Krohn:2009wm,Bhattacherjee:2012ir,Tweedie:2014yda,Godbole:2019erb}. In \cite{Kaplan:2008ie,Krohn:2009wm,Tweedie:2014yda,Godbole:2019erb}, interesting techniques and several interesting observables pertaining to top polarization measurement have been proposed. In~\cite{Bhattacherjee:2012ir}, the polarization of hadronic top from the stop decay has been probed in the context of 13 TeV LHC. For all these studies involving hadronic top, boosted jet substructure~\cite{Ellis:2009su,Ellis:2009me,Kitadono:2014hna,Kitadono:2015nxf,Conway:2016caq,Lapsien:2016zor,Kogler:2018hem} has been employed for extraction of polarization information. In ~\cite{Bhattacharya:2020vzu}, machine-learning techniques have been used for boosetd top-tagging as well as polarization measurement. Furthermore, top polarization has been studied also using the matrix element method~\cite{Brandenburg:2002xr,Baumgart:2012ay,Tweedie:2014yda}.

In this work, we probe top polarization from stop decay in the semileptonic decay mode at high luminosity and high energy LHC. For this purpose, we have considered all the relevant backgrounds and various kinematic observables to reduce the background and find a suitable signal region, where the purported signal can be observed with reasonable significance. We have chosen such signal benchmarks, which are allowed by all the experimental constraints and also can be probed at the early runs of high-luminosity and high energy LHC. For polarization measurement, we have utilized angular observables proposed in ~\cite{Godbole:2019erb} and was shown to be effective in the LHC environment. Lastly, we have performed a statistical analysis and calculated the significance with which the left- and right-handed top quark from the stop decay can be separated from each other at the future runs of LHC. In this context, we have also performed a comparative study of different centre-of-mass energies and integrated luminosities.

The plan of this paper is as follows. In Section~\ref{sec2}, we discuss the phenomenology of stop-mixing in MSSM and resulting polarization of top quark from the stop decay. In Section~\ref{sec3}, we discuss the signal processes and all the relevant backgrounds. In Section~\ref{sec4}, we present the simulation and analysis strategy to separate signal and background at high-luminosity and high energy LHC. We analyse the polarization observable and compute asymmetries for our signal benchmarks in the presence of background events in Section~\ref{sec5}. The statistical analysis and calculation of $P$-value or corresponding significance level for discriminating left- and right-polarized tops from stop decay are done in Section~\ref{sec6}. In Section~\ref{sec7} we summarize and conclude our analysis.

\section{Polarization of top from stop decay} 
\label{sec2}

The heaviest known fermion, top quark has a special property compared to all other known quarks. It decays before hadronization. Top polarization can be measured looking at its decay products, since the decay products retain the information of the top quark spin. The kinematics of top decay is extremely interesting and has been studied in detail~\cite{Kane:1991bg,Hisano:2003qu,Shelton:2008nq,Perelstein:2008zt,Arunprasath:2016tfq}. Top decays to $b$ quark and $W$ boson with almost 100\% branching ratio (BR). The $W$-boson further decays to hadronic(BR 70\%) or leptonic (BR 30\%) final states. The angular distribution of any of the decay products in the top rest frame is given by the following relation~\cite{Jezabek:1988ja,Czarnecki:1990pe}. 

\begin{equation}
\frac{1}{\Gamma}\frac{d\Gamma}{d\cos\theta_f} = \frac{1}{2}(1+{\cal P}_0\kappa_f\cos\theta_f) 
\label{pol}
\end{equation}    

\noindent
Where $\theta_f$ is the angle between the fermion momentum and top quark spin direction in the top rest frame. In this paper, we work with helicity basis, ie. the top quark spin-quantization axis is same as the top quark direction of motion. Nevertheless, Equation~\ref{pol} will be valid irrespective of the top spin-quantization axis. ${\cal P}_0$ is the top polarization~\cite{Shelton:2008nq,Mahlon:1999gz,Schwienhorst:2010je,Arunprasath:2016tfq} in the corresponding basis, whose value lies between -1 and +1. $\kappa_f$ is called the spin-analyzing power. $\kappa_f$ can be calculated~\cite{Jezabek:1994qs} and its values in SM at the tree-level are as follows~\footnote{For hadronic top decay the correction to the tree-level value of spin-analyzing power can be 3-4\%~\cite{Brandenburg:2002xr,Bernreuther:2014dla}}.

\begin{eqnarray}
\kappa_{\bar d} = \kappa_{\ell} = 1,~~~ \kappa_{u} = -0.3,~~~ \kappa_b = -0.4
\label{spin}
\end{eqnarray} 

\noindent
It is evident from Equations~\ref{pol} and \ref{spin} that $\bar d$/$\ell$ have the maximal correlation with top spin. However, $b$-quark or correspondingly the reconstructed $W$ boson (with spin analysing power equal and opposite of that of $b$ quark) can also be used for studying top polarization.

In this work, we focus on the polarized top from the decay of top squark in MSSM. One of the major decay modes of the lightest stop is $\tilde{t_1} \rightarrow t \tilde{\chi}_1^0$. The branching ratio in this channel vary depending on the masses of various sparticles. Here $\tilde{\chi}_1^0$ is the lightest neutralino and also the lightest supersymmetric particle(LSP), which will give rise to considerable $\slashed{E_T}$ in the final state. The chiral structure of the top quark interactions in this case can be understood from the relevant part of the Lagrangian.

\begin{equation}
{\cal L} = \bar{\tilde{\chi}}_1^0(g^{\tilde{t}_{1L}}P_L + g^{\tilde{t}_{1R}}P_R)t\tilde{t}_1
\end{equation}

\noindent
Where $P_L=\frac{1-\gamma_5}{2}$ and $P_R=\frac{1+\gamma_5}{2}$ are the the projection operators and $g^{\tilde{t}_{1L}}$ and $g^{\tilde{t}_{1R}}$, are the couplings between top quark, stop and neutralino, which are responsible for the polarization of top from the stop decay. The form of these couplings are given below. 

\begin{equation}
g^{\tilde{t}_{1L}}  =  -\sqrt{2}g_{2}
\left[\frac{1}{2}Z_{12}^{*}+\frac{1}{6}\tan\left(\theta_{W}\right)Z_{11}^{*}\right]
\cos\theta_{\tilde t}
-\left[\frac{g_{2}m_{{t}}Z_{14}^{*}}{\sqrt{2}m_{W}\sin\left(\beta\right)}\right]
\sin\theta_{\tilde t}
\label{lag1}
\end{equation}
\begin{equation}
g^{\tilde{t}_{1R}}  = 
\left[\frac{2\sqrt{2}}{3}g_{2}\tan\left(\theta_{W} \right)Z_{11}\right]
\sin\theta_{\tilde t}-\frac{g_{2}m_{t}Z_{14}}{\sqrt{2}m_{W}\sin\left(\beta\right)}
\cos\theta_{\tilde t}
\label{lag2}
\end{equation}
\noindent
The parameters in Equation~\ref{lag1} and \ref{lag2} are the mixing elements in the neutralino sector as well as in the stop sector, given as follows. 

\begin{eqnarray}
\label{mixing}
    \tilde{\chi}_1^0 & = & Z_{11}\tilde{B}+Z_{12}\tilde{W}_{3}+Z_{13}\tilde{H}_{d}+Z_{14}\tilde{H}_{u}\\
    \tilde{t}_{1} & = & \tilde{t}_{L} \cos{\theta_{\tilde{t}}}
    + {\tilde{t}_{R}} \sin{\theta_{\tilde{t}}}
\end{eqnarray}

\noindent
$Z_{ij}$s are mixing elements in the neutralino sector, whereas $\theta$ is the mixing angle between $\tilde{t}_L$ and $\tilde{t}_R$. It is clear from Equation~\ref{lag1},~\ref{lag2} and ~\ref{mixing} that the couplings receive contribution from both gauge($\tilde{B},\tilde{W}_{3}$) and Higgs($\tilde{H}_{u},\tilde{H}_{d}$) sector. The gauge interactions conserve chirality and the Yukawa interactions flip it. The stop mixing and gaugino and Higgsino component in the lightest neutralino will essentially determine the polarization of the top quark produced from the decay of a heavy stop. The resulting conditions are given as follows:

\begin{itemize}
\item Pure gaugino-like $\tilde{\chi}^0_1$ and $\cos \theta_{\tilde{t}} = 0$ the top produced is right-handed.
\item Pure gaugino-like $\tilde{\chi}^0_1$ and $\sin \theta_{\tilde{t}} = 0$ the top produced is left-handed.
\item Pure Higgsino-like $\tilde{\chi}^0_1$ and $\cos \theta_{\tilde{t}} = 0$ the top produced is left-handed.
\item Pure Higgsino-like $\tilde{\chi}^0_1$ and $\sin \theta_{\tilde{t}} = 0$ the top produced is right-handed.
\end{itemize}

\noindent 
Evaluating top polarization via Equation~\ref{pol}, relies on correct reconstruction of the top rest frame. In case of hadronically decaying top, it is possible to reconstruct the top quark fully and top rest frame can be identified. But in case of leptonically decaying top, the full reconstruction of the top rest frame is not possible, more so when there are additional sources of $\slashed{E_T}$ over and above that from the top decay (often the case when top comes from the decay of heavy states linked with some NP). It has been proposed~\cite{Perelstein:2008zt}, to construct an approximate top rest frame in case of leptonic decay of top quark, while it is produced in association with another top quark decaying hadronically. In \cite{Shelton:2008nq}, measurement of top polarization in the leptonic final state, in terms of energy observables in the lab frame, have been explored in highly boosted regime($\beta \rightarrow 1$). In~\cite{PrasathV:2014omf}, a further generalized analysis has been performed, where top polarization have been probed, in the leptonic decay mode, in terms of both angular and energy observables in the lab frame, for varied boost of the top quark. 

Furthermore, hadronic decay of top quark has much higher yield in terms of number of events, compared to the leptonic decay. One should also note that, the isolated lepton identification becomes difficult in the boosted regime as the lepton merges with the hadronic decay products, when the top quark is highly boosted~\cite{Rehermann:2010vq}. We will discuss this issue in detail in the subsequent sections. Measuring top polarization, in case of boosted hadronically decaying tops have been studied in \cite{Bhattacherjee:2012ir,Godbole:2019erb}. 

In this work, we focus on a scenario where top quark is produced via decay of TeV scale stop, because those will be most accessible from the stand-point of discovery at the current and future LHC. We will concentrate on the semileptonic decay mode of a top quark pair from the decay of a pair of heavy stops for the following reasons. Firstly, the semileptonic mode has significant signal yield ($\approx$ 44\% BR, comparable to the fully hadronic mode with BR $\approx 45\%$)~\cite{Zyla:2020zbs}. In addition, It is relatively a cleaner channel, compared to the fully hadronic decay mode, due to the presence of an isolated lepton in the final state. On the other hand, the presence of one hadronic top helps us identifying the top rest frame correctly, in contrast to the fully leptonic decay mode.

\section{Signal and Backgrounds at the LHC}
\label{sec3}

We present our analysis in the context of future runs of LHC. We will focus on a few benchmark cases of proposed runs, namely high luminosity LHC (HL-LHC)~\cite{ATLAS:2018zrp} with 14 TeV centre-of-mass energy, high energy LHC(HE-LHC)~\cite{FCC:2018bvk} with 33 TeV centre-of-mass energy and FCC-hh~\cite{FCC:2018vvp} with 100 TeV run. 
Events for the signals and the corresponding backgrounds have been generated using Madgraph@MCNLO~\cite{Alwall:2014hca} and their cross-sections have been calculated at the next-to-leading order(NLO). We have used the nn23lo1 parton distribution function. MLM matching with xqcut = 30 GeV is performed for backgrounds with multiple jets in the final state. PYTHIA8~\cite{Sjostrand:2006za} has been used for the showering and hadronization and the detector simulation has been taken care of by Delphes-3.4.1~\cite{deFavereau:2013fsa}. Jets are formed by the built-in Fastjet~\cite{Cacciari:2006sm} of Delphes.

We are interested in exploring the reach of high luminosity. In order to reach an integrated luminosity of ($\sim 3000 fb^{-1}$), it will be imperative to increase the instantaneous luminosity and that will increase the number of pile-up events per bunch-crossing significantly compared to earlier LHC runs. We follow the analysis of ~\cite{Cohen:2013xda}, and assume 140 average pile-up interactions. Pile-up subtraction algorithm of Delphes~\cite{deFavereau:2013fsa} has been employed in this case.

{\bf Signal} :
We are interested in the boosted regime of top, where top is produced from the decay of a heavy stop at the LHC. Therefore, the decay products of top will be largely collimated. We focus on the case where one top is decaying hadronically and the other leptonically. The hadronic decay product of top will form a fatjet in this case. Hence we are interested in a final state which contains at least 1 top-tagged fatjet + 1 isolated lepton + at least 1 $b$-tagged jet + $\slashed{E_T}$. 

The decay chain we are focussing on is the following:

$p p \rightarrow \tilde{t}_1 \tilde{t}_1 \rightarrow t\tilde{\chi}^0_1\bar t\tilde{\chi}^0_1 \rightarrow bW^+\bar bW^-\tilde{\chi}^0_1\tilde{\chi}^0_1 \rightarrow$ top-tagged fatjet + $b \ell \nu_{\ell}$

\subsection{Choice of signal benchmarks}

\begin{figure}[!hptb]
	\centering
	\includegraphics[width=10cm,height=9cm]{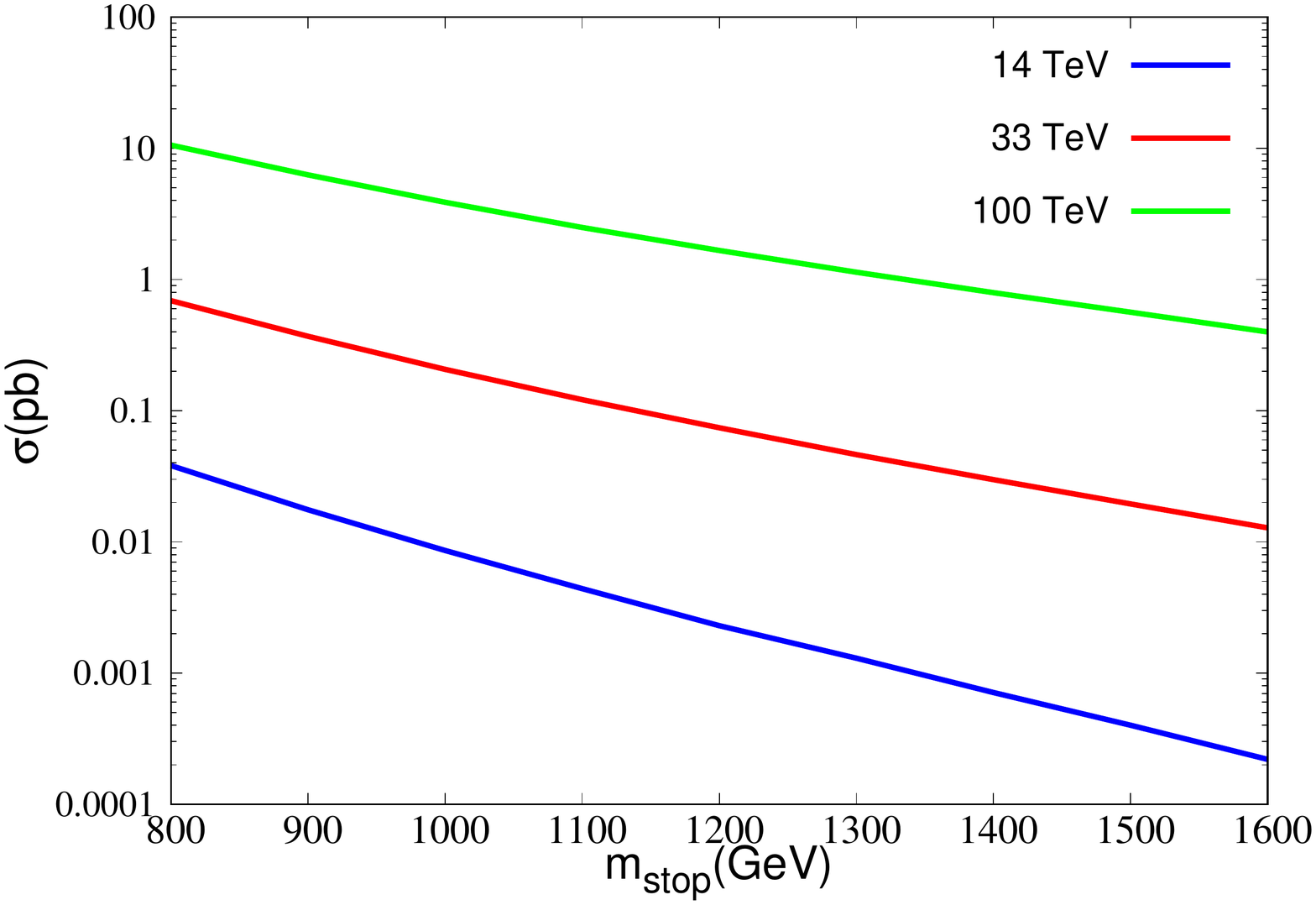}
	\caption{Production cross-section of stop pair at the LHC at 14 TeV, 33 TeV and 100 TeV centre-of-mass energy, as a function of stop mass.}
	\label{stop_crosssec}
\end{figure}

Our objective in this work is to distinguish between hadronically decaying left and right-handed top quarks from the stop decay at the LHC. The polarization of the stop is determined by the composition of the stop and neutralino. From the stand-point of detectability of the signal at the LHC, the mass of ${\tilde{t}_1}$ will play a crucial role. In Figure~\ref{stop_crosssec}, we show the how the stop pair production cross-section varies with stop mass. On the other hand, the subsequent decay chain of ${\tilde{t}_1}$, which leads to our purported signal depends on the branching ratio of ${\tilde{t}} \rightarrow t \tilde{\chi}^0_1$. This branching fraction depends on (a) the Higgsino-gaugino fraction in the lightest neutralino, (b) mass of the lightest neutralino mass (c) mass of other sparticle states that can arise from the decay of stop. When the lightest neutralino, is gaugino-like and all other neutralinos and charginos are heavier the the stop, the stop decays almost exclusively to the decay mode ${\tilde{t}} \rightarrow t \tilde{\chi}^0_1$. If the lightest neutralino is Higgsino-like, the stop decays into ${\tilde{t}} \rightarrow t \tilde{\chi}^0_1$ channel, with 50\% branching ratio. Because, in case of Higgsino-like lightest neutralino, the lightest chargino is also degenerate with the lightest neutralino opening another decay mode for stop, ${\tilde{t}} \rightarrow b \tilde{\chi}^{\pm}_1$ with 50\% branching ratio. While choosing our benchmarks, we will consider both of these scenarios. The set of mass for and ${\tilde{t}}, \tilde{\chi}^0_1$ is actually restricted by the experimental limit coming from the direct search for stop in the following decay modes ${\tilde{t}} \rightarrow t \tilde{\chi}^0_1$ or ${\tilde{t}} \rightarrow b \tilde{\chi}^{\pm}_1$~\cite{CMS:2019ysk,CMS:2021qkg}. 

\noindent
Keeping all these aspects in mind, we choose our benchmarks, as follows in Table~\ref{bps}. We mention here that, we have chosen such benchmarks in \ref{bps}, which differ in production rate at LHC.

\begin{table}[!hptb]
\begin{center}
\begin{tabular}{| c | c | c | c | c |}
\hline
Benchmarks  & $m_{\tilde{t}}$ & $m_{\tilde{\chi^0_1}}$ & $Z_{11}$ & $Z_{14}$ \\
\hline
BP1 & 1.4 TeV & 100 GeV & 1 & 0 \\
\hline
BP2 & 1.4 TeV & 100 GeV & 0 & 1 \\
\hline
BP3 & 900 GeV & 700 GeV & 1 & 0 \\
\hline
BP4 & 900 GeV & 700 GeV & 0 & 1 \\
\hline
\end{tabular}
\caption{List of signal benchmark points}
\label{bps}
\end{center}
\end{table}

\noindent
We see in Table~\ref{bps}, that BP1 and BP3 correspond to gaugino-like LSP, whereas BP2 and BP4 correspond to Higgsino-like LSP. This property will not only govern their production rate, it also plays a crucial role in determining the polarization of the top quark coming from the stop decay. In Table~\ref{sig_crosssec}, we give the corresponding production cross-section$\times$branching ratio, in the specific channel of our interest at various centre of mass energies. BP3 and BP4 have significantly higher production rate compared to BP1 and BP2. On the other hand, in BP3 and BP4, stop mass and the mass gap between stop and neutralino mass are smaller compared to that of BP1 and BP2. Therefore, in BP3 and BP4, the produced top quark will have lower boost factor. With decreasing boost-factor, the fatjet analysis in the merged category becomes less-suited and one should use resolve category analysis~\cite{CMS:2019ysk}. However, we will use merged category analysis for all the benchmarks, to investigate the reach of the fatjet analysis in terms of top boost in the context of top tagging as well as polarization measurement in realistic LHC environment.

 \begin{table}[!hptb]
\begin{center}
\begin{tabular}{| c | c | c | c |}
\hline
Benchmarks & 14 TeV & 33 TeV & 100 TeV \\
\hline
BP1 & 0.31 fb & 13.2 fb & 352 fb\\
\hline
BP2 & 0.16 fb & 6.6 fb & 176 fb\\
\hline
BP3 & 7.8 fb & 162 fb & 2723 fb \\
\hline
BP4 & 3.9 fb & 81.2 fb & 1361 fb \\
\hline
\end{tabular}
\caption{Production cross-sections(NLO) of signal benchmarks at various $\sqrt{s}$.}
\label{sig_crosssec}
\end{center}
\end{table}

\subsection{List of backgrounds}
The background processes contributing to the aforementioned final state are as follows:

\begin{itemize}
\item $t \bar t$ semileptonic: The dominant background for our final state is the SM $t \bar t$ production where, one of the tops decays hadronically and the other leptonically, giving rise to the a top-tagged fatjet + one isolated lepton + $b$-jet + $\slashed{E_T}$. The source of $\slashed{E_T}$ here is the neutrino from $W$ decay. 
\item $t \bar t~\text{semileptonic}+ Z(Z\rightarrow\nu_{\ell}\bar \nu_{\ell})$: This background is irreducible just like the $t \bar t$ semileptonic background, when one of the tops decays hadronically and the other leptonically, and $Z$ boson decays into a pair of neutrinos, further adding to the source of $\slashed{E_T}$. This background, owing to large $\slashed{E_T}$, contribute substantially to our signal region, despite having much smaller cross-section.  
\item $t \bar t$ leptonic: Another important background comes from $t \bar t$ production, when both the tops decay leptonically. This background can be important when one of the leptons has either escaped detection, due to finite detector acceptance or it has failed the isolated lepton identification criteria. However, only a meager fraction of this background can give rise to a reconstructed hadronic tagged top, which is only possible through mistagging.
\item $W(W\rightarrow\ell\nu_{\ell}$)+jets: The next important background is $W$+jets, where $W$ decays leptonically. Here too, the source of $\slashed{E_T}$ is $W$ boson. Although in this final state, there is no $b$-tagged jet and top jet can only occur at the detector due to mistagging. However, the large cross-section of this process makes it inevitable to consider this background with due care.
\item $WZ(Z\rightarrow\nu_{\ell}\bar \nu_{\ell}$): A subdominant background comes from the $WZ$ production in SM, when the $Z$ boson decays into a pair of neutrinos. We have considered both the possibilities for $W$ decay. When $W$ decays leptonically, there is an isolated charged lepton in the final state and the $b$-jet and top fatjet are coming from mistagging. On the other hand, when $W$-decays hadronically, which happens with higher probability, a light-jet can be mistagged as lepton.
 
\end{itemize}

In Table~\ref{bkgcrosssec}, we show the cross-sections of all the aforementioned background processes at various centre of mass energies of our interest.

\begin{table}[!hptb]
\begin{center}
\begin{tabular}{| c | c | c | c |}
\hline
Background  & 14 TeV & 33 TeV & 100 TeV \\
\hline
$t \bar t$ semileptonic & 358 pb & 958 pb & 6629 pb  \\
\hline
$W + $ jets & 2.6 $\times 10^4$ pb & 5.8 $\times 10^4$ pb & 1.6 $\times 10^5$ pb \\
\hline
$t \bar t$ leptonic & 42 pb & 162 pb & 1106 pb \\
\hline
$WZ$ & 2.0 pb & 6.5 pb  & 25.3 pb \\
\hline
$t \bar t$(semileptonic)+$Z$ & 0.038 pb & 0.28 pb & 2.4 pb \\
\hline
\end{tabular}
\caption{Production cross-sections(NLO) of background processes at various $\sqrt{s}$.}
\label{bkgcrosssec}
\end{center}
\end{table}

\section{Simulation and Analysis}
\label{sec4}
\subsection{Kinematic distributions}

We first present the distributions of kinematic observables, that can play crucial role in signal background discrimination. We would like to mention here that the kinematic distributions for BP1 and BP2 are expected to be same, since the two benchmarks only differ in terms of gaugino-Higgsino mixing parameter, affecting only the corresponding branching ratios in those channels. Similar statement is true for BP3 and BP4. Therefore, we present the kinematic distributions for BP1 and BP3. We would like to compare the distributions between the left-and right handed cases pertaining to BP1 and BP3 as well. For that purpose, we make use of the mixing angle in the stop sector. From our earlier discussion we know that $Z_{11} = 1, Z_{14} = 0, \sin\theta = 0$ will correspond to left handed top, which we call BP1 left-handed top. Similarly, $Z_{11} = 1, Z_{14} = 0, \cos\theta = 0$ will lead to right-handed top, and will be discussed as BP1 right-handed case. Exactly in same way, BP3 left- and right-handed cases are defined.

\begin{figure}[!hptb]
	\centering
         \includegraphics[width=7cm,height=6cm]{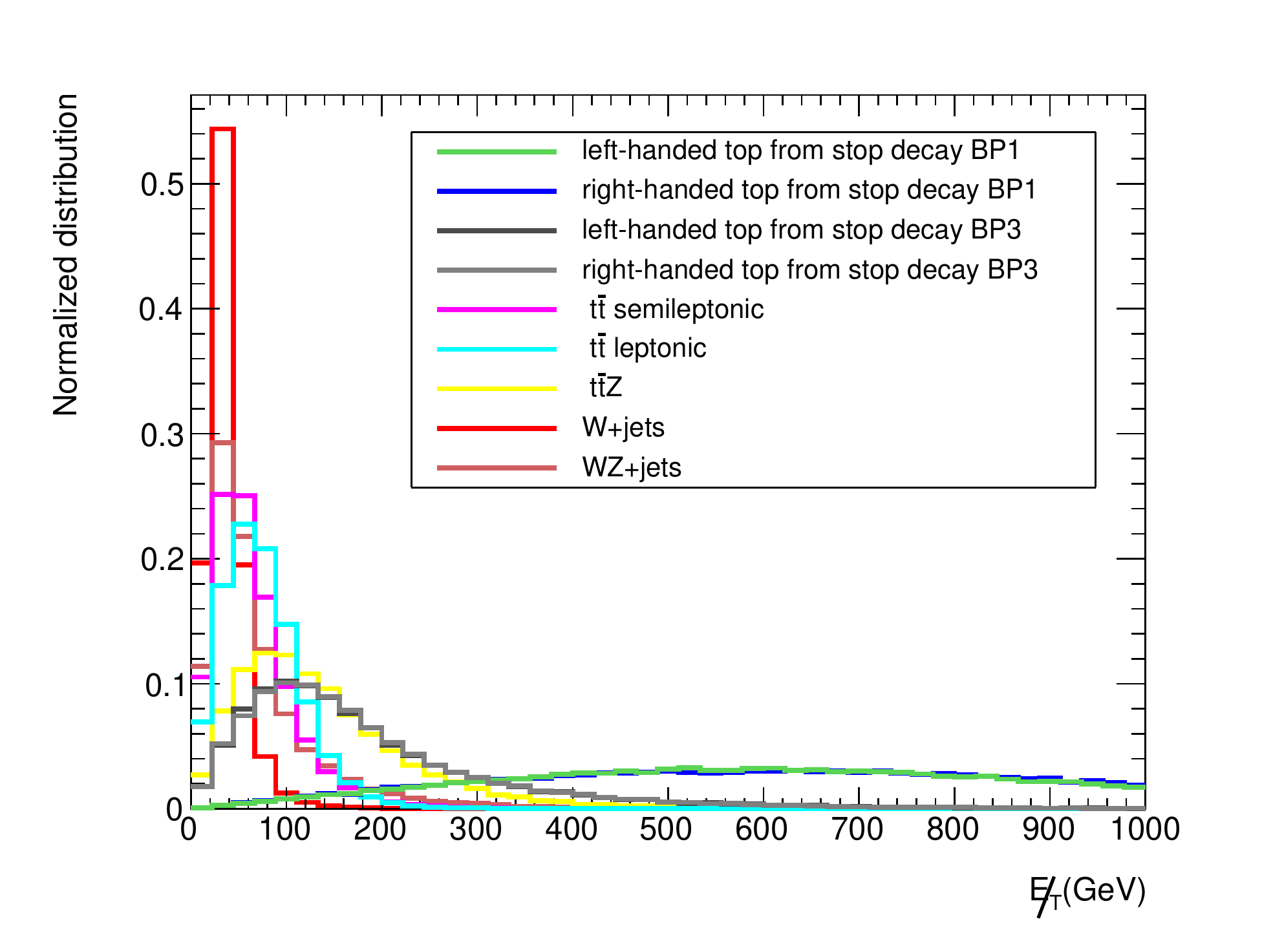}
\includegraphics[width=7cm,height=6cm]{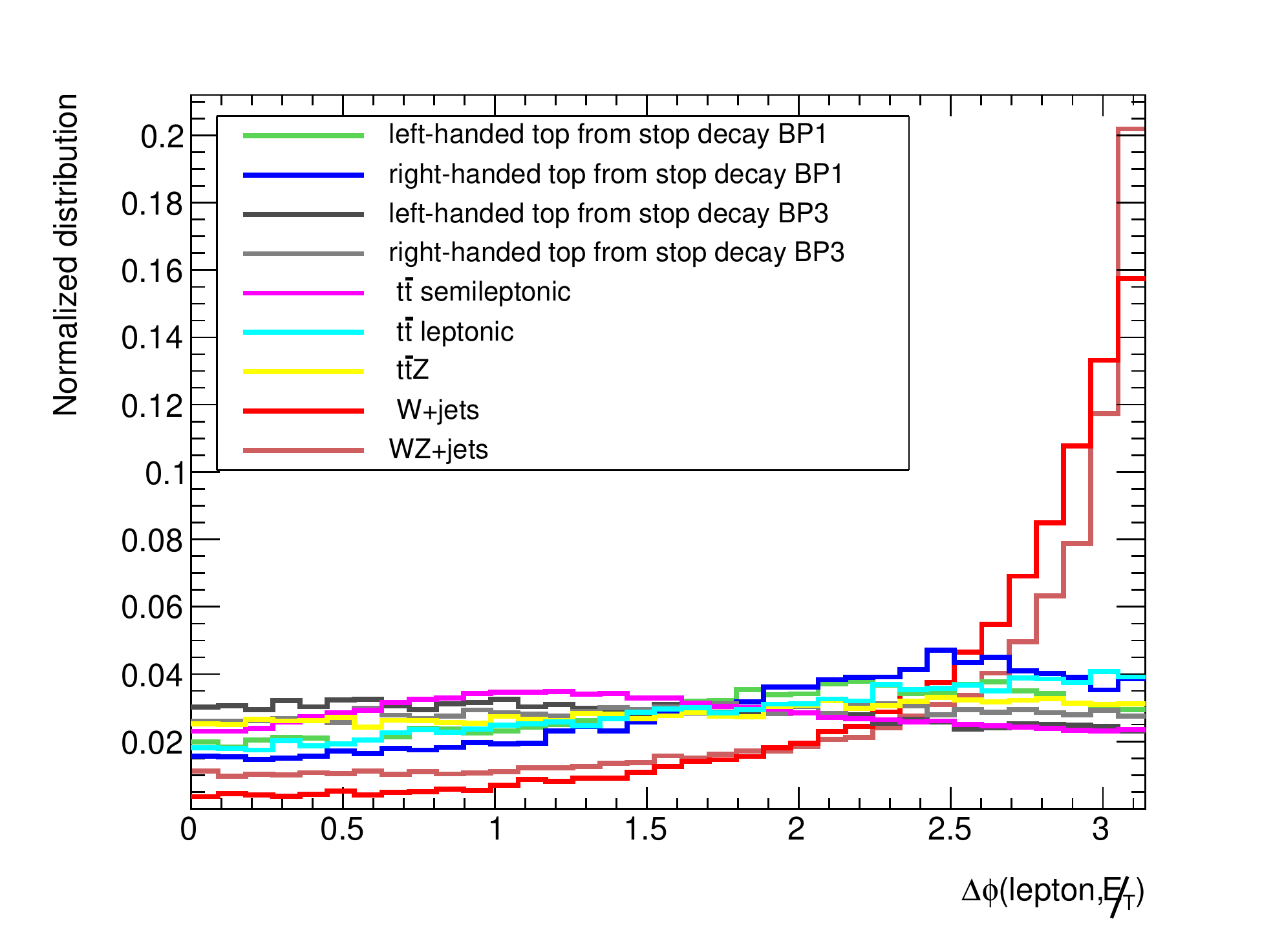} \\
  \includegraphics[width=7cm,height=6cm]{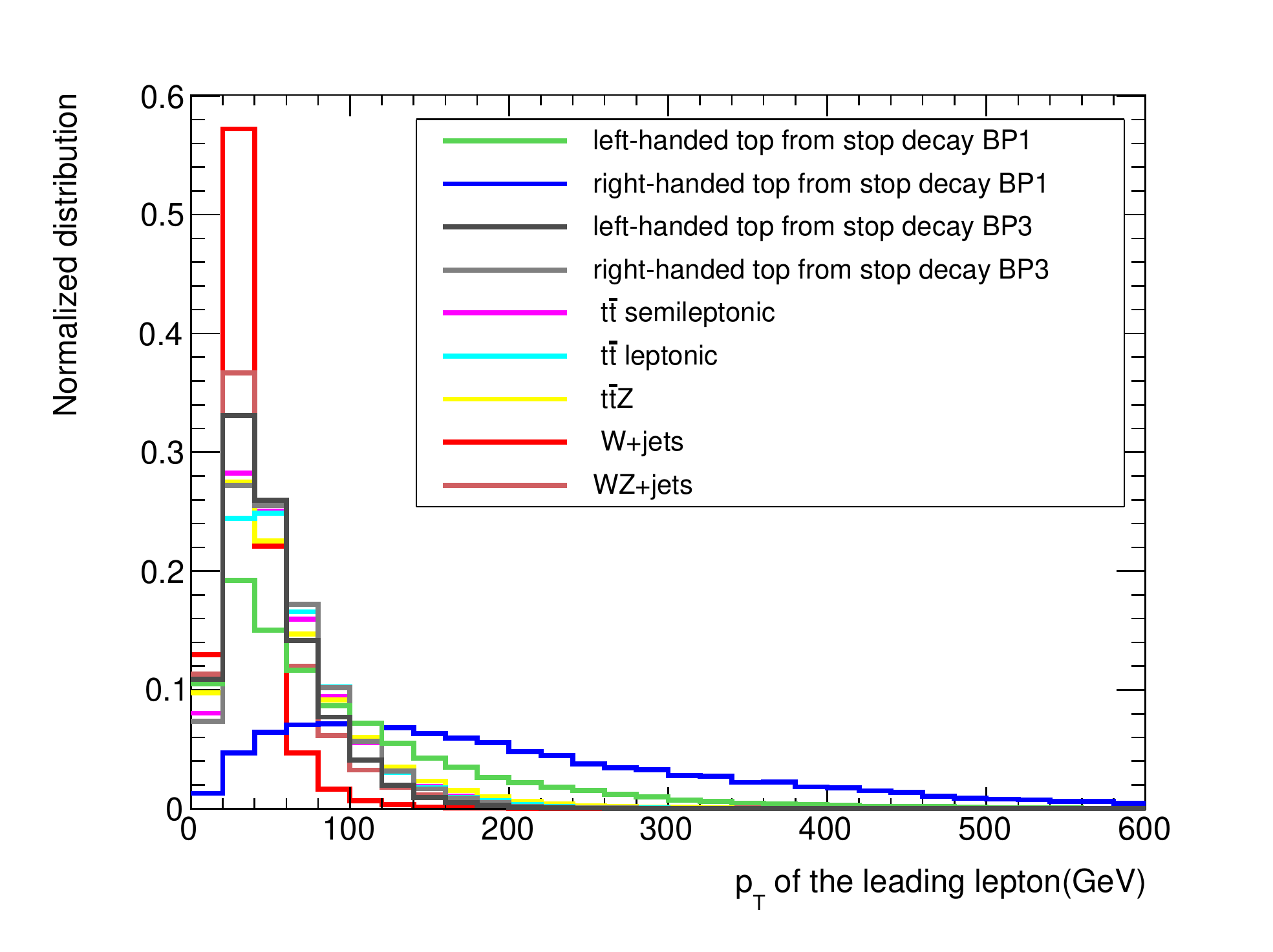}
         \includegraphics[width=7cm,height=6cm]{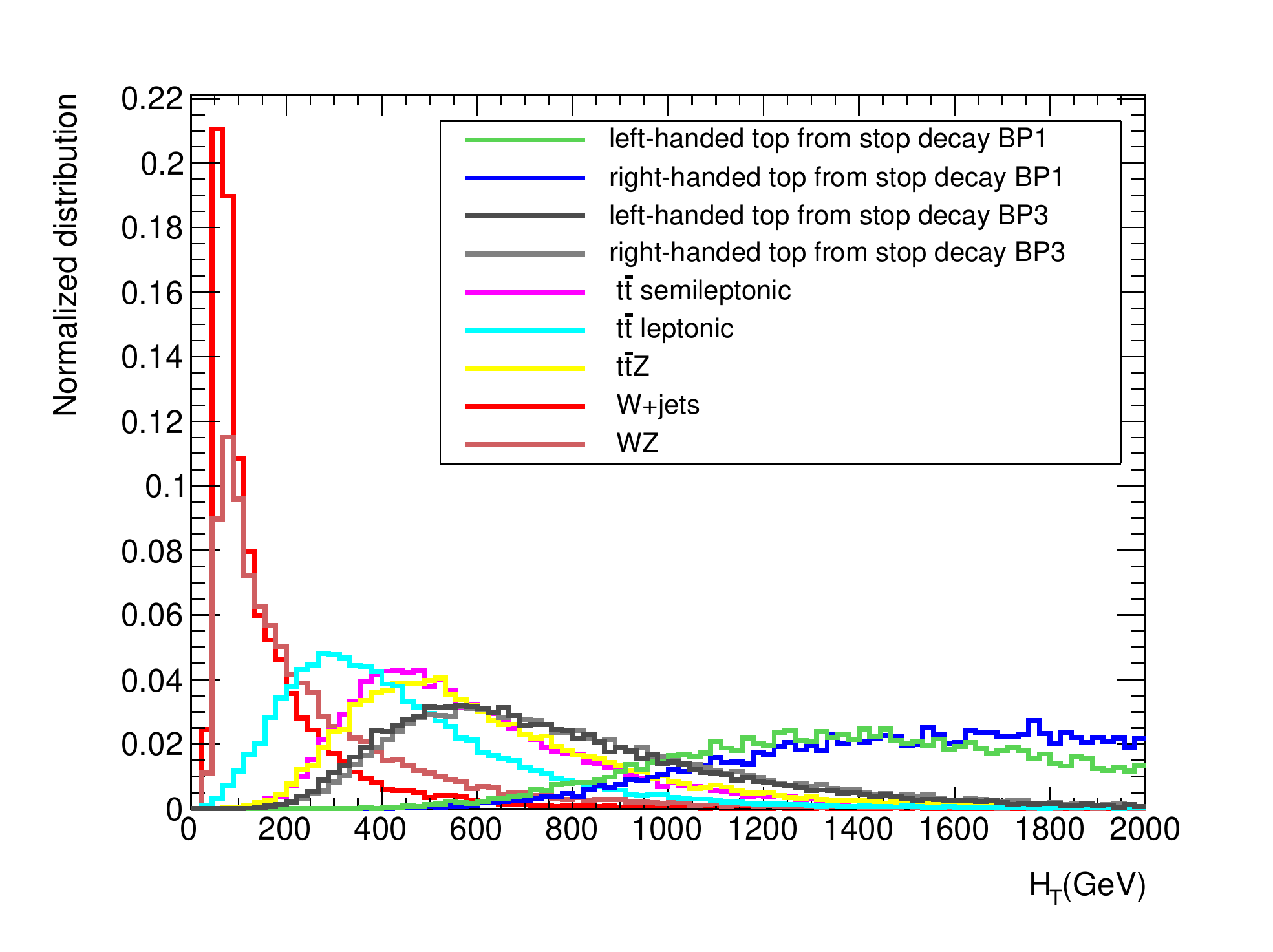} \\
\includegraphics[width=7cm,height=6cm]{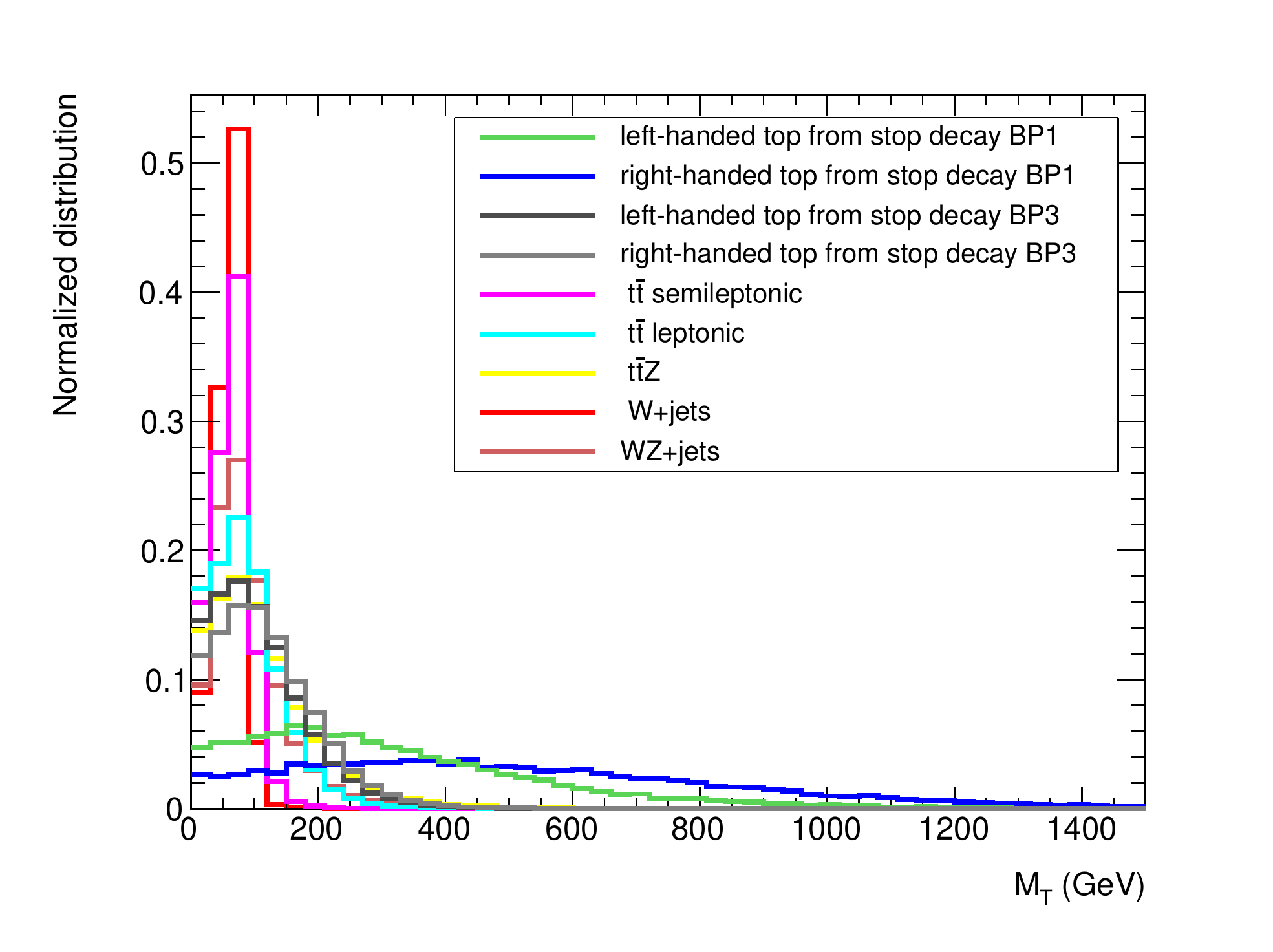} 
  \includegraphics[width=7cm,height=6cm]{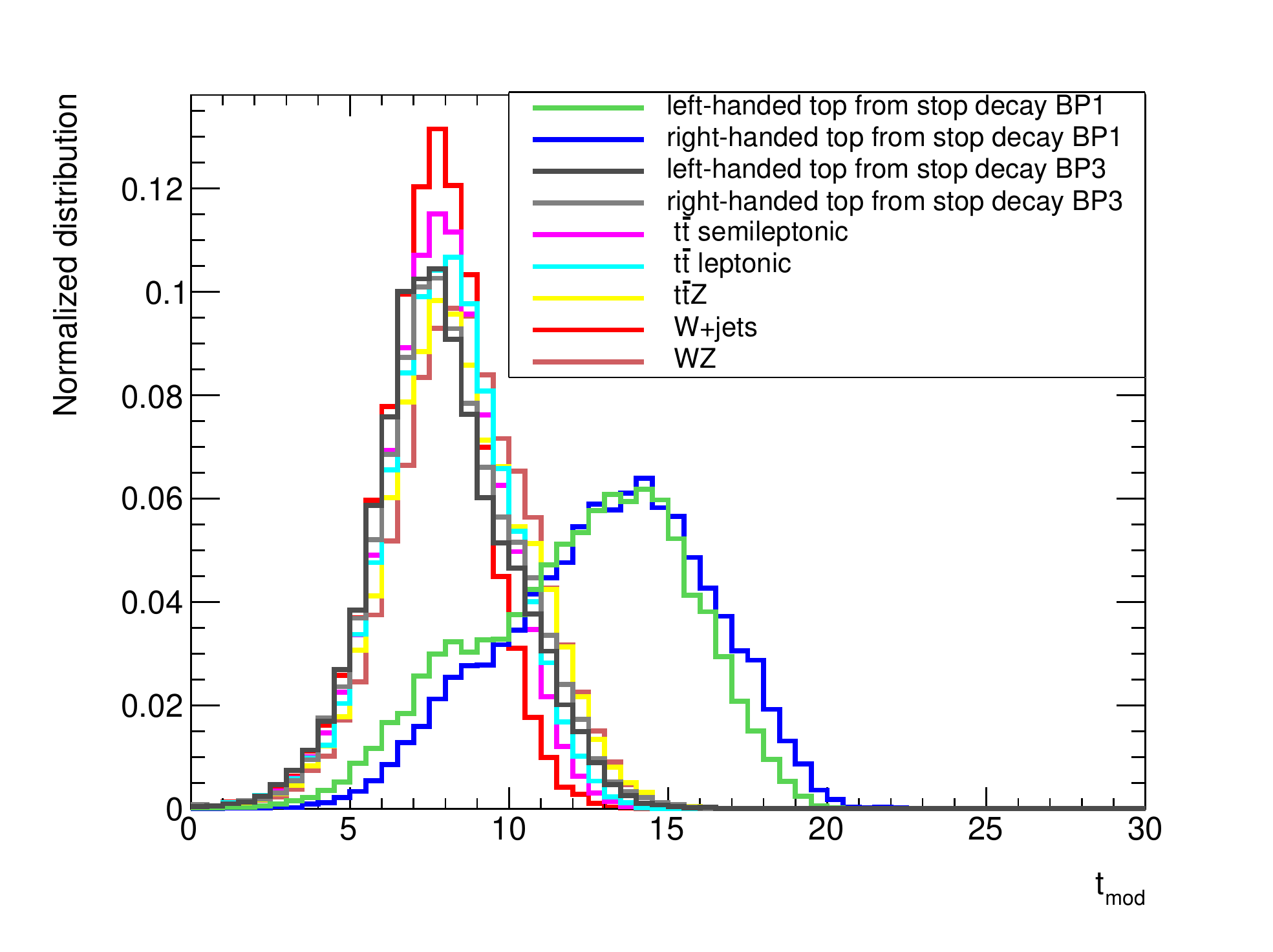}
	\caption{Normalized distribution for signal as well as backgrounds.}
	\label{dist}
\end{figure}

{\begin{itemize}
\item $\slashed{E_T}$: First, we present the $\slashed{E_T}$ distribution. In case of signal, the source of $\slashed{E_T}$ is the neutralino pair production from the heavy stop decay. A large mass gap between the stop and neutralino will ensure large $\slashed{E_T}$ in the signal. In case of BP1, this mass gap is larger compared to that of BP3. Therefore, $\slashed{E_T}$ distribution peaks at significantly higher value in case of BP1 than in BP3. On the other hand, for almost all the background processes, the $\slashed{E_T}$ comes from neutrino coming from $W$ decay. Therefore the $\slashed{E_T}$ distribution peaks at a much lower value in case of backgrounds, which we can see from Figure~\ref{dist}(top left).

\item $\Delta \phi(\vec{\ell},\vec{\slashed{E_T}})$ : We can see from Figure~\ref{dist}(top right) that, the in case of $W$+jets background this distribution peaks at a large value. This happens because in case of $W$+jets background, the lepton and $\slashed{E_T}$ comes from the decay of $W$ and are almost back to back, making large azimuthal angle between the lepton and $\slashed{E_T}$. In case signal and $t \bar t$ backgrounds that is not the case.

\item $p_T^{\ell}$ :  The $p_T$ distribution of the leading lepton is shown in Figure~\ref{dist}(centre left). One can see that the $p_T$ distribution of the leading lepton peaks at higher value for signal processes, since the lepton comes from the decay of a heavy stop. However, one can also see that the lepton $p_T$ distribution for left-handed top peaks at a lower value compared to that of the right-handed top. The reason behind is the angular correlation between the lepton and $b$-jet from top decay which is sensitive to the top polarization. In case of left-handed top $b$-jet carries large fraction of the top energy and the lepton carries a smaller fraction. As a result, the $p_T$ distribution of the leading lepton peaks at a smaller value in case of left-handed top, compared to the right-handed case.

\item $H_T$ :  We show the $H_T$ distribution in Figure~\ref{dist}(centre right). $H_T$ is defined in this case as the scalar sum of $p_T$ of all visible final state particles. The figure shows that the signal benchmarks, especially BP1 produces much larger $H_T$ compared to the backgrounds. BP3, although does not produce large $H_T$, still $H_T$ distribution provides some amount of signal-background separation, as we will see in the cut flow in the next section.

\item $M_T$ : $M_T$ or transverse mass~\cite{Tovey:2010de} has played a very important role in SUSY searches~\cite{CMS:2019ysk}. The transverse mass variable is defined as follows:

\begin{equation}
M_T = \sqrt{2 \slashed{E_T} |\vec{p_T}^{\ell}| (1-\cos(\Delta \phi(\vec{\ell},\vec{\slashed{E_T}})))}
\end{equation}

One can easily check that $M_T^2 \leq M_W^2$, when the $\slashed{E_T}$ is coming from a single neutrino coming from $W$-decay. Therefore, in such cases the end-point of the $M_T$ distribution will correspond to $W$ mass. For example, in case of $t \bar t$ semileptonic and $W$+jets background, $M_T$ distribution indeed shows this feature. However, for signal the $\slashed{E_T}$ is not coming from a single $W$ and therefore $M_T$ peaks at a much larger value. However, the end-point for signal depends on the mass of the parent particle. BP1 shows a larger $M_T$ spread as compared to BP3 for this reason. We can see this from Figure~\ref{dist}(bottom left). We also see from the same figure that $M_T$ takes higher values for right-handed top compared to the left-handed tops. It can be easily understood from the aforementioned lepton $p_T$ distribution. Larger lepton $p_T$ in case of right-handed top pushes $M_T$ to a larger value. It is worth mentioning that $t \bar t$ leptonic and $t \bar tZ$ backgrounds will not obey the $M_T$ cut-off, because they consist of additional sources of $\slashed{E_T}$ other than that from $W$-decay. 

\item Modified topness ($t_{mod}$) : To reduce $t \bar t$ leptonic background(where one lepton is missed) one needs to construct an observable called modified topness or $t_{mod}$ which has been introduced in ~\cite{Graesser:2012qy} and has been used in the context of reducing dileptonic $t \bar t$ background effectively~\cite{CMS:2016sth,CMS:2019ysk}. $t_{mod}$ is defined as follows.

\begin{equation}
t_{mod} = ln(min S) ~~~\text{where} ~~ S = \frac{(m_W^2 - (p_{\nu} + p_{\ell})^2)^2}{a_W^4} + \frac{(m_t^2 - (p_b + p_W)^2)^2}{a_t^4}
\label{tmod}
\end{equation}

$a_W = 5$GeV and $a_t=15$GeV are resolution parameters. $t_{mod}$ is essentially a $\chi^2$-like variable which distinguishes between signal and $t \bar t$-leptonic events. The first term in Equation~\ref{tmod} corresponds to the decay of the particular top in which the lepton is identified and reconstructed. The second term corresponds to the top decay corresponding to the ``lost" lepton. The quantity $p_W$ is basically the sum of four-momenta of the ``lost" lepton and neutrino (since both of them contribute to the $\slashed{E_T}$). The minimization is done with respect to the free parameters, all three-momenta of $\vec{p_W}$, and the component of $\vec{p_{\nu}}$ along the beam-axis, with the constraints $\vec{\slashed{E_T}} = \vec{p_W} + \vec{p_{\nu}}$ and $p_W^2 = m_W^2$. It is clear that for the leptonic top background this observable will peak at a much lower value compared to that of the signal and can be used for reduction of this background in particular. The distribution of $t_{mod}$ for signal and backgrounds can be seen in Figure~\ref{dist}(bottom right)

\end{itemize}

\subsection{Top-tagging}

A crucial part of our analysis is the tagging of boosted hadronic tops. It is likely that the decay products of the top quark will be merged as a single fatjet in the highly boosted regime. We then look for substructures inside such a top-fatjet and from the substructure try to reconstruct the parent top quark. However, this process is quite non-trivial and requires a detailed algorithm involving jet grooming. There are multiple top-taggers available and widely in use such as JHTagger~\cite{Kaplan:2008ie}, CMSTopTagger~\cite{CMS:2014fya,CMS:2019gpd}, HEPTopTagger~\cite{Plehn:2010st,Kasieczka:2015jma} etc. In our work, we have done our analysis using HEPTopTagger. The HEPToptagger algorithm has been validated using experimental data~\cite{ATLAS:2016rgb}. We present the top tagging efficiency and top-mistag rate(probability that a light-jet is tagged as top fatjet), for various range of $p_T$ of the fatjet of our interest, in Table~\ref{roc}.

 \begin{table}[!hptb]
\begin{center}
\begin{tabular}{| c | c | c |}
\hline
$p_T$ range in GeV & top-tagging efficiency & mistag rate \\
\hline
200-300 & 3.1\% & 0.1\% \\
\hline
300-400 & 15\% & 0.6\% \\
\hline
400-500 & 29\% &  1.4\% \\
\hline
500-600 & 35\% & 2\% \\
\hline
600-700 & 43\% & 2.47\% \\
\hline
700-800 & 46\% & 2.44\% \\
\hline
\end{tabular}
\caption{Top-tagging efficiency and top-mistag rate as a function of fatjet $p_T$.}
\label{roc}
\end{center}
\end{table}

 We have used SM $t\bar t$ sample for calculating top-tagging efficiency and QCD dijet events for calculating the top-mistag rate. Our results are in agreement with the tagging efficiency and mistag rates obtained by the experimental groups~\cite{CMS:2011xsa,CMS:2016tvk}. The following criteria have been applied for fatjet grooming.

\begin{itemize}
\item We have formed fatjets using Cambridge-Achen(CA) algorithm with radius $\Delta R \approx 1.0$.
\item The maximum subjet mass is 30 GeV. 
\item Mass drop~\cite{Butterworth:2008iy} threshold is taken to be 0.8.
\item The filtering~\cite{Plehn:2011tg} radius $R_{filt} = 0.3$
\item Number of subjets after filtering $n_{filt} = 5$ 
\item The minimum $p_T$ of the filtered subjets are required to be 30 GeV.
\end{itemize}

\medskip
\noindent
After the jet-grooming is done, the following criteria are applied for top-tagging.

\begin{itemize}
\item The minimum $p_T$ of the top candidate is taken to be 200 GeV.
\item The top mass reconstruction criterion is applied on the top fatjet candidate ie. 150 GeV $ < m_{topjet} < 200$ GeV.
\item The $W$ reconstruction with the fatjet is done with $\Delta m < 0.15 M_W$.
\end{itemize}

\begin{figure}[!hptb]
	\centering
	\includegraphics[width=9cm,height=7cm]{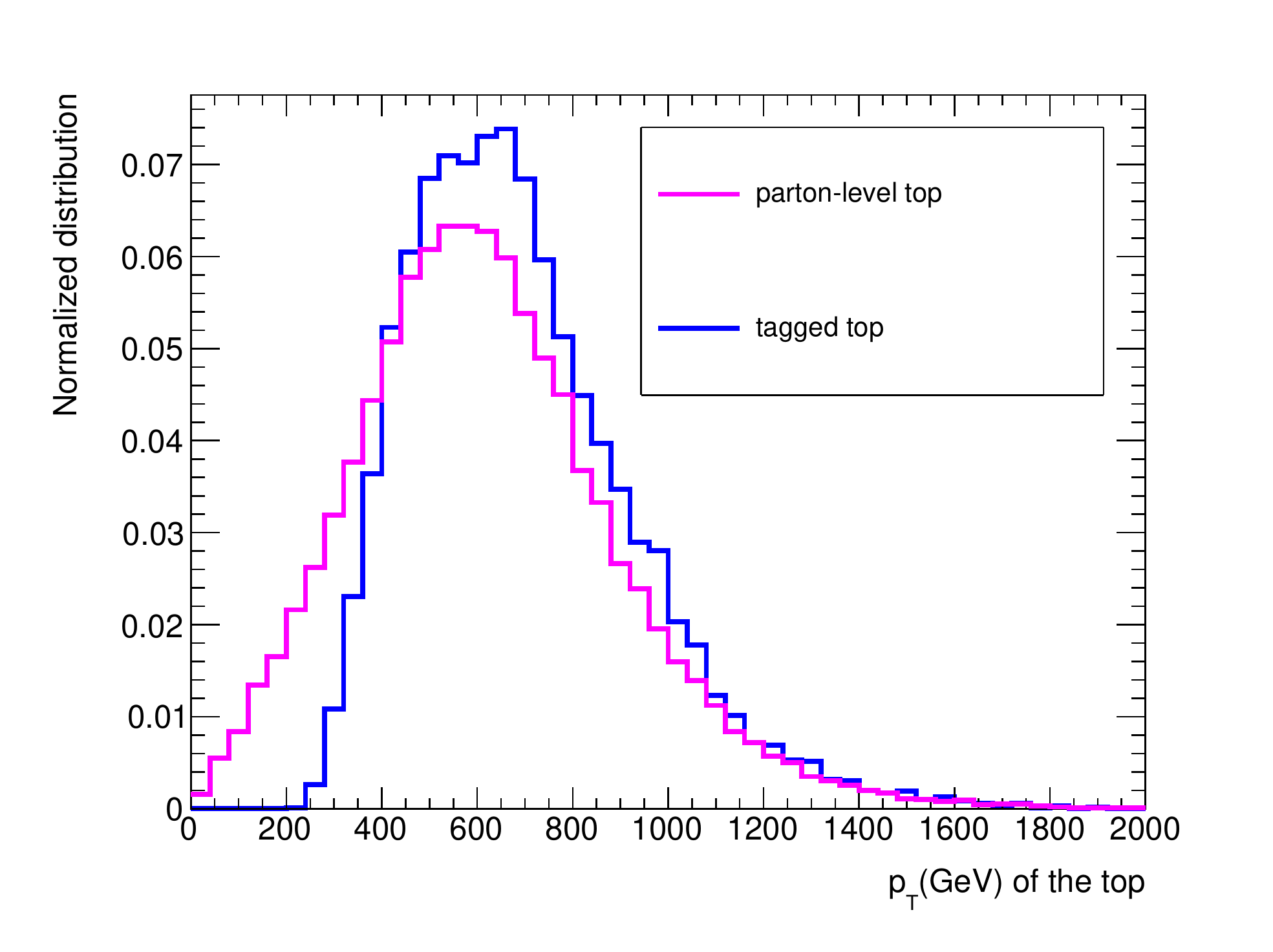}
	\caption{Normalized distribution of parton-level and tagged top(BP1).}
	\label{toppt}
\end{figure}


\noindent
In Figure~\ref{toppt}, we show the comparison between the $p_T$ distribution of parton-level top and tagged top for BP1 for demonstrating the performance of the top-tagger in terms of top-$p_T$ reconstruction. It is clear that the $p_T$ distribution of the tagged top is slightly shifted to a higher value compared to the parton-level top. The reason behind this is the tagging efficiency for low $p_T$ tops is quite small, and only a small fraction of low $p_T$ tops will fall under the boosted and merged category and will be tagged by the algorithm adopted by HEPTopTagger.  However, at high-$p_T$ regime, the $p_T$ reconstruction is quite accurate for the same reason described above.

\begin{figure}[!hptb]
	\centering
	\includegraphics[width=9cm,height=7cm]{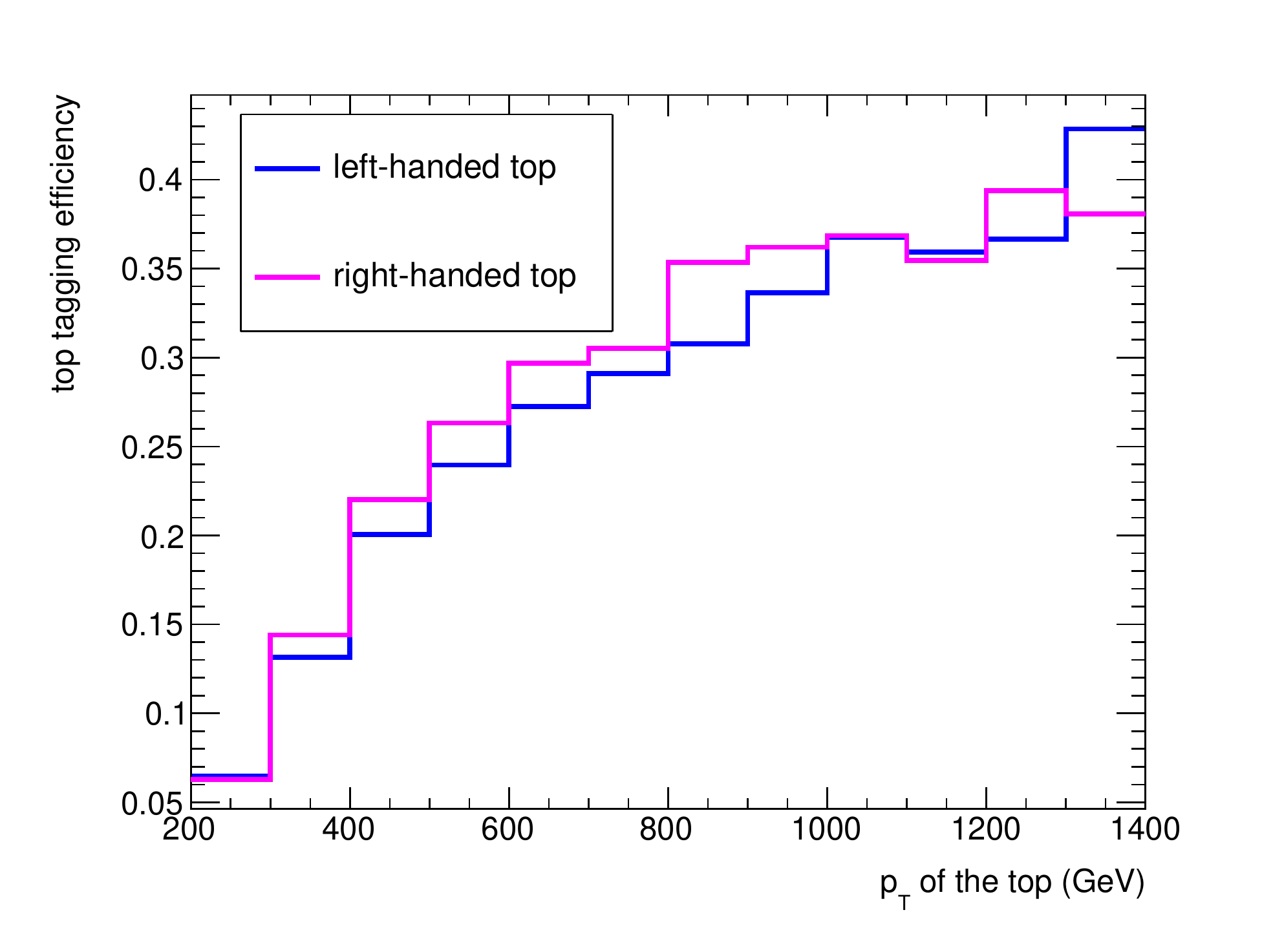}
       \caption{Top tagging efficiency as a function of $p_T$ of the parton-level top for left and right-handed case(BP1).}
	\label{efficiency}
\end{figure}

In Figure~\ref{efficiency}, we show the top-tagging efficiency as a function of $p_T$ of the parton-level top quark for left and right-polarized cases of BP1. We can see that the tagging efficiency is actually very small when the $p_T$ of the top quark is low. The tagging efficiency increases with increase in top $p_T$. This is because the in case of large boost, the decay products of top quark will be confined in a large $R$ fatjet. But in case of low boost, they will be largely separated and it will not be possible to tag the top using jet-substructure analysis with HEPTopTagger. It is also seen from Figure~\ref{efficiency}, that the tagging efficiency is slightly better in the right-handed case than in left-handed case. In left-handed top, the $b$-subjet energy fraction is much larger compared to that of the right-handed case. Therefore the energy fraction carried by the $W$-boson is smaller and the $W$-reconstruction criterion is not satisfied in some cases, which explains the slightly lower tagging efficiency of the left-handed tops.
 
\begin{figure}[!hptb]
	\centering
	\includegraphics[width=9cm,height=7cm]{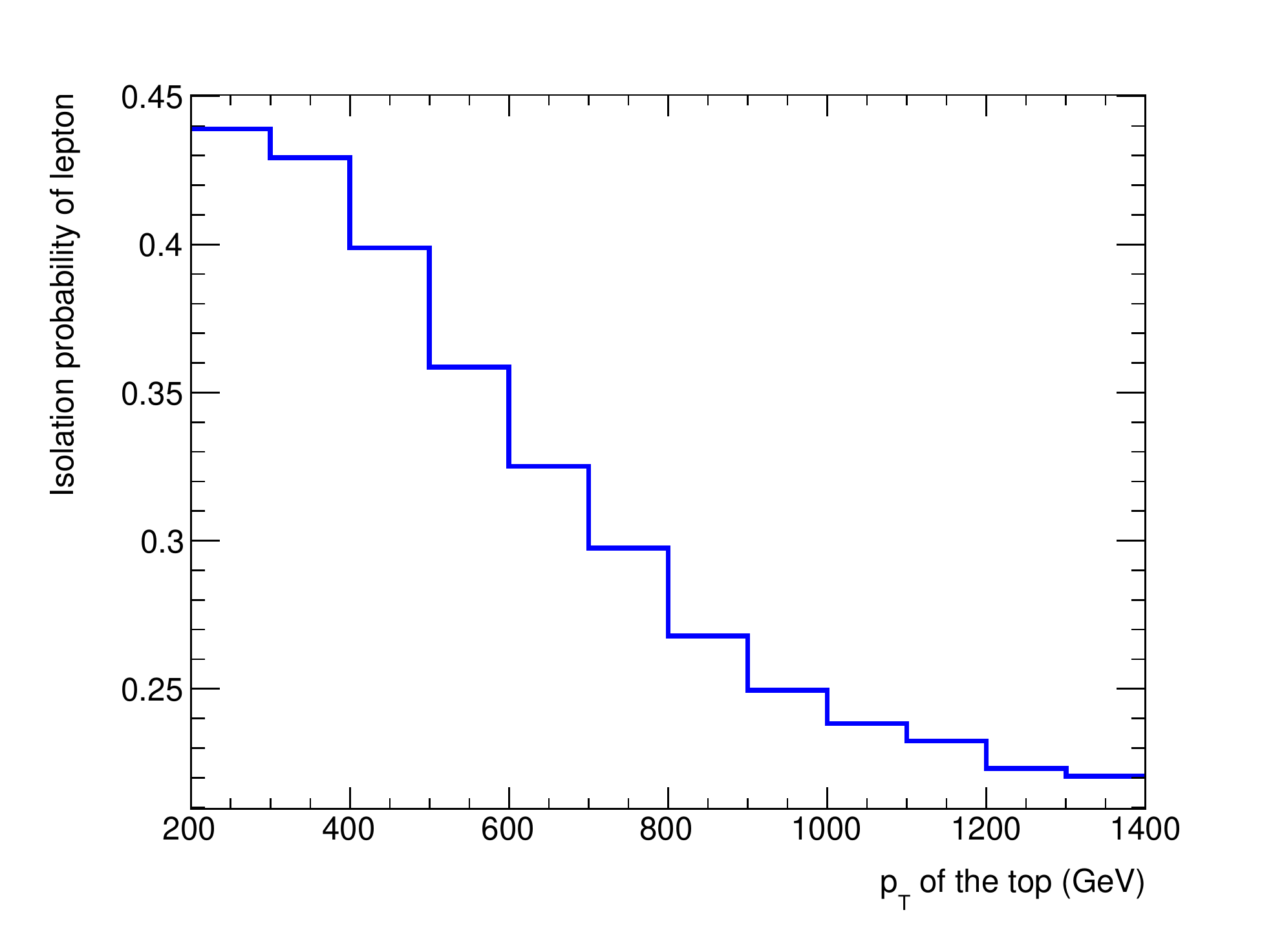}
       \caption{Probability of identifying an isolated lepton as a function of $p_T$ of the parton-level top for the lepton isolation cone radius $\Delta R \sim 0.5$.}
	\label{lepton_iso}
\end{figure}

In case of semileptonic decay of top quark, one top is hadronically decaying forming a fatjet and the other top is decaying leptonically, leading to a lepton and a $b$-jet in the final state. In this case, a crucial selection criterion becomes demanding one isolated lepton in the final state. In case of a boosted top, the lepton and $b$-jet coming from the top will be very close to each other, and their energies will not be very different. It is pointed out in ~\cite{Rehermann:2010vq} that, the opening angle between the lepton and $b$-jet is inversely proportional to the $p_T$ of the top-quark, ie. $\Delta R \sim \frac{m_t}{p_T^{t}}$. In BP1/BP2, the $p_T^t \approx 600$ GeV and therefore $\Delta R \sim 0.3$. 
In such cases, the lepton will fail to satisfy the standard isolation criteria (which Delphes uses with CMS-specifications) given below.
\begin{equation}
\frac{\sum_{i} p_T^i(\Delta R < 0.5)}{p_T^{\gamma}} < 0.12
\end{equation}

\noindent
In Figure~\ref{lepton_iso}, we actually see this feature clearly. We see that as the top $p_T$ increases, the probability of getting an isolated lepton in the event decreases.  In other words, in typical signal events, the advantage gained in terms of top-tagging efficiency in the high $p_T$ region is dwarfed by the decrease in lepton isolation criteria in the same $p_T$ range. The semileptonic final state pertaining to the process is actually faced with this challenge. However, as it was pointed out in \cite{Rehermann:2010vq}, one can use $p_T$ scaled $\Delta R$ to solve this issue. We have used a modified $\Delta R \sim 0.3$ in our case, which ensures 55-60\% isolation probability for leptons throughout the entire top $p_T$ range of our interest.

\begin{figure}[!hptb]
	\centering
	\includegraphics[width=7cm,height=6cm]{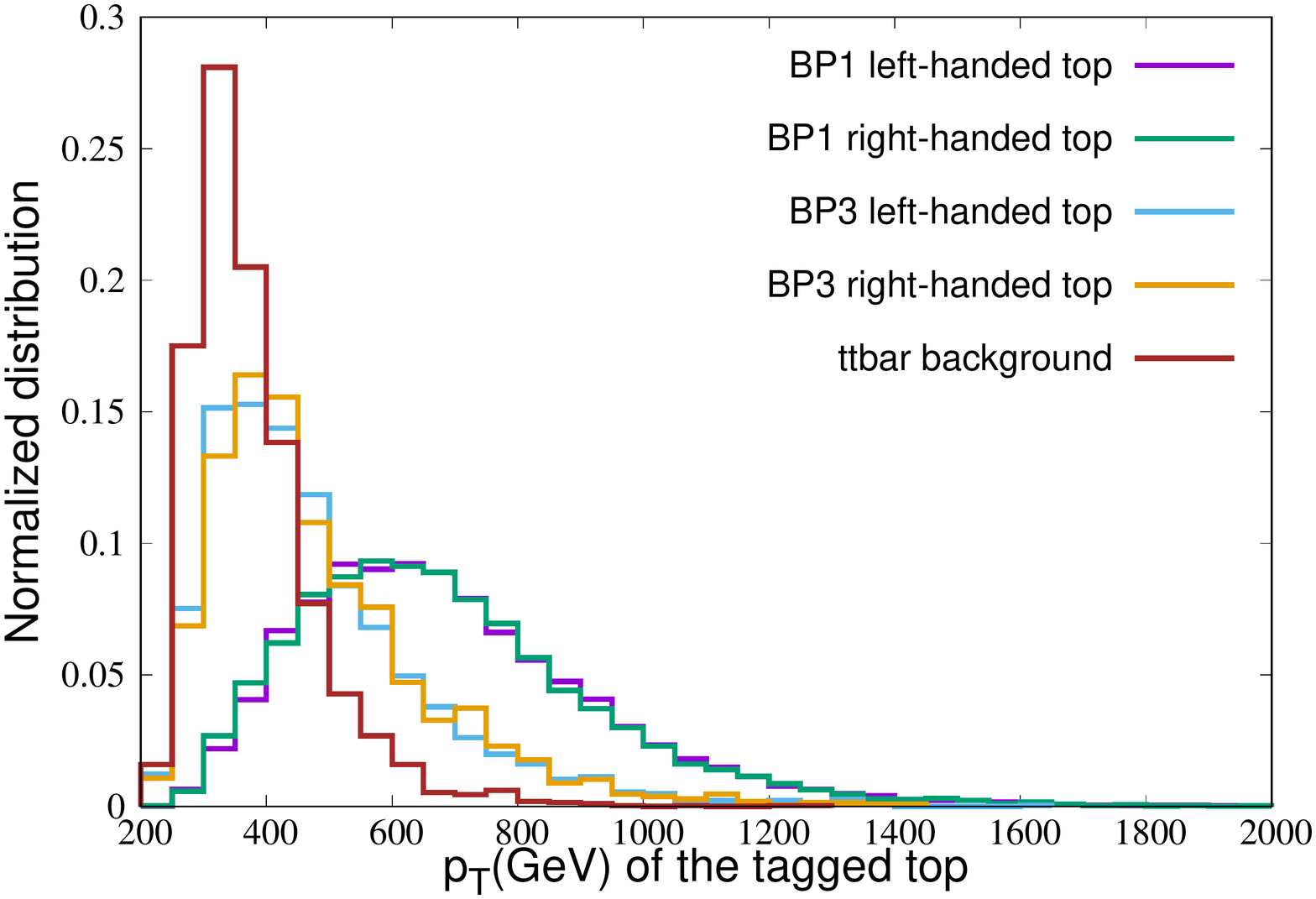} 
        \includegraphics[width=7cm,height=6cm]{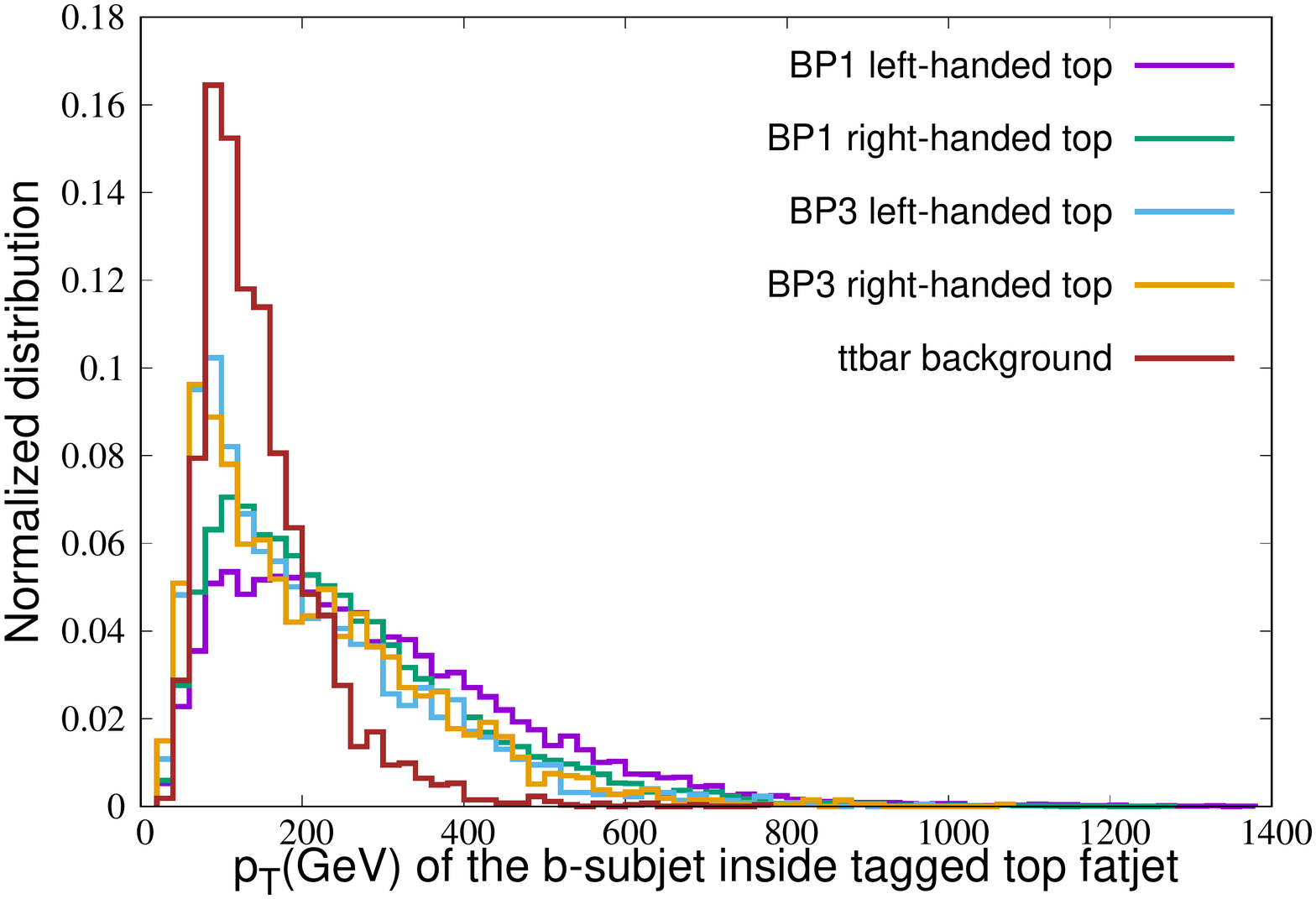}
	\caption{Normalized distribution for $p_T$ of the reconstructed tagged-top(left) and the $p_T$ of the reconstructed $b$-subjet(right) for left- and right-handed top from stop decay as well as $t \bar t$ semileptonic background.}
	\label{tagged_dist}
\end{figure}

In Figure~\ref{tagged_dist},(left) we plot the distribution of $p_T$ of the tagged top for left and right-handed tops from stop decay for BP1 and BP3 and the major background pertaining to semileptonic $t \bar t$. Clearly in BP1, the $p_T$ distribution of top from decay of heavy stop and in association with a light neutralino peaks at a large value. The $p_T$ distribution for BP3 as well as $t \bar t$ background, peaks at a much lower value, as expected. We also show the $p_T$ distribution of the reconstructed $b$-subjet, inside the top fatjet in Figure~\ref{tagged_dist}(right). It is clear that $p_T$ of the $b$-subjet is larger in case of signal benchmarks. It is also seen that in case of left-handed top the $p_T$ distribution of $b$-subjet is somewhat shifted to higher values, which is again related to the angular correlation between top decay products.

\subsection{Results}

Having discussed the relevant features of the kinematic observables for signal and background processes, we find out a suitable cut-flow which will 
increase our signal vs background separation to a desired degree.

\noindent
{\bf Cut flow:}

\noindent
Pre-selection criteria : One isolated lepton and at least one $b-$jet \\
Cut1 : $\slashed{E_T} >$ 300 GeV \\
Cut2 : $M_T > 150$ GeV \\
Cut3 : $\Delta \phi(\vec{\ell},\vec{\slashed{E_T}}) < 2.9$ \\
Cut4 : modified top mass $t_{mod} > 10$ \\
Cut5 : $H_T > 1000$ GeV \\
Final selection criteria : At least one fatjet tagged as top fatjet

\begin{table}[!hptb]
\scriptsize{
\begin{tabular}{| c | c | c | c | c | c | c | c |}
\hline
Datasets & selection cuts & Cut1 & Cut2 & Cut3 & Cut4 & Cut5 & top-tagged \\
\hline
BP1 left-handed & 437 & 372 & 307 &  266 & 232  & 214  & 58 \\
\hline
BP1 right-handed & 409 & 353 & 324 & 276  &  247 & 238  & 70 \\
\hline
BP2 left-handed & 218 & 186 & 154 &  133 & 116  & 107  & 29 \\
\hline
BP2 right-handed & 205 & 177 & 162 & 138 & 123  & 119  &  35  \\
\hline
BP3 left-handed & 14110 & 2200 & 1100 & 983 & 351  &  257 &  21  \\
\hline
BP3 right-handed & 14157 & 2200 & 1264 & 1147  & 421  &  328 & 24 \\
\hline
BP4 left-handed & 7055 & 1100 & 550 & 492  &  176 & 129 &  11 \\
\hline
BP4 right-handed & 7079 & 1100 & 632 & 574  & 211  &  164 & 12 \\
\hline
$t \bar t$ semileptonic & 5.37$\times 10^8$  & 805500  & 3759 & 3759 & 537  & 537  & 537 \\
\hline
$W$+jets & 5.4$\times 10^8$ & 446400 & 26640 &  19728  & 12960  & 12960  &  65 \\
\hline
$t \bar t$ leptonic & 4.5$\times 10^7$ & 50400 & 20160 & 12600  & 5796  & 2898  & 84 \\
\hline
$t \bar t Z$ & 57000 & 3192 & 1938 & 1596  &  1117 & 502  & 100 \\
\hline
$WZ$ & 114000 & 6000 & 2520 &  2220 &  1380 & 660  & 60 \\
\hline
\end{tabular}
\caption{Signal and background events surviving after applying various cuts and selection criteria at $\sqrt s = 14$ TeV and ${\cal L} = 3000 fb^{-1}$.}
\label{tablecutflow14}
}
\end{table}

\begin{table}[!hptb]
\scriptsize{
\begin{tabular}{| c | c | c | c | c | c | c | c |}
\hline
Datasets & selection cuts & Cut1 & Cut2 & Cut3 & Cut4 & Cut5 & top-tagged \\
\hline
BP1 left-handed & 6205 & 5282 & 4359 &  3777 & 3294  & 3039  & 824 \\
\hline
BP1 right-handed & 5808 & 5013 & 4601 & 3919  &  3507 & 3380  & 994 \\
\hline
BP2 left-handed & 3096 & 2641 & 2187 &  1889 & 1647  & 1519  & 412 \\
\hline
BP2 right-handed & 2911 & 2513 & 2300 & 1960 & 1747  & 1690  &  497  \\
\hline
BP3 left-handed & 98770 & 15400 & 7700 & 6881 & 2457  &  1799 &  147  \\
\hline
BP3 right-handed & 99099 & 15400 & 8848 & 8029  & 2947  &  2296 & 168 \\
\hline
BP4 left-handed & 49385 & 7700 & 3850 & 3444  &  1232 & 903 &  77 \\
\hline
BP4 right-handed & 49553 & 7700 & 4424 & 4018  & 1477  &  1148 & 84 \\
\hline
$t \bar t$ semileptonic & 4.7$\times 10^8$  & 718500  & 3353 & 3353 & 479  & 479  & 479 \\
\hline
$W$+jets & 4.3$\times 10^8$ & 3.7$\times 10^5$ & 15892 &  15892  & 10440  & 10440  &  52 \\
\hline
$t \bar t$ leptonic & 5.7$\times 10^7$ & 64800 & 25920 & 16200  & 7452  & 3726  & 108 \\
\hline
$t \bar t Z$ & 140000 & 7840 & 4760 & 3920  &  372 & 2744  & 246 \\
\hline
$WZ$ & 123500 & 6500 & 2730 &  2405 &  1495 & 715  & 65 \\
\hline
\end{tabular}
\caption{Signal and background events surviving after applying various cuts and selection criteria at $\sqrt s = 33$ TeV and ${\cal L} = 1000 fb^{-1}$.}
\label{tablecutflow33}
}
\end{table}

\begin{table}[!hptb]
\scriptsize{
\begin{tabular}{| c | c | c | c | c | c | c | c |}
\hline
Datasets & selection cuts & Cut1 & Cut2 & Cut3 & Cut4 & Cut5 & top-tagged \\
\hline
BP1 left-handed & 16540 & 14080 & 11620 &  10068 & 8781  & 8100  & 2192 \\
\hline
BP1 right-handed & 15480 & 13361 & 12263 & 10446  &  9349 & 9008  & 2649 \\
\hline
BP2 left-handed & 8251 & 7040 & 5829 &  5034 & 4390  & 4050  & 1097 \\
\hline
BP2 right-handed & 7759 & 6699 & 6132 & 5223 & 4655  & 4504  &  1325  \\
\hline
BP3 left-handed & 164195 & 25601 & 12800 & 11439 & 4084  &  2991 &  244  \\
\hline
BP3 right-handed & 164742 & 25601 & 14701 & 13347  & 4899  &  3817 & 279 \\
\hline
BP4 left-handed & 82097 & 12800 & 6400 & 5725  &  2048 & 1501 &  128 \\
\hline
BP4 right-handed & 82377 & 12800 & 7354 & 6679  & 2455  &  1908 & 140 \\
\hline
$t \bar t$ semileptonic & 3.3$\times 10^8$  & 497175  & 2320 & 2320 & 331  & 331  & 331 \\
\hline
$W$+jets & 1.2$\times 10^8$ & 99200 & 5920 &  4384  & 2880  & 2880  &  14 \\
\hline
$t \bar t$ leptonic & 3.9$\times 10^7$ & 44240 & 17696 & 11060  & 5087  & 2543  & 74 \\
\hline
$t \bar t Z$ & 120000 & 6720 & 4080 & 3360  &  2352 & 1057  & 211 \\
\hline
$WZ$ & 48070 & 2530 & 1063 &  936 &  582 & 278  & 25 \\
\hline
\end{tabular}
\caption{Signal and background events surviving after applying various cuts and selection criteria at $\sqrt s = 100$ TeV and ${\cal L} = 100 fb^{-1}$.}
\label{tablecutflow100}
}
\end{table}

We can see from Table~\ref{tablecutflow14},~\ref{tablecutflow33} and \ref{tablecutflow100}, that our chosen variables and the cut-flow reduce the background events to a large extent. As we have already discussed the $\slashed{E_T}$ and $M_{T}$ plays major role in reducing the $t \bar t$ semileptonic background. In case of $t \bar t$ leptonic background $\slashed{E_T}$ and $M_T$ prove to be little less effective because it has additional sources of $\slashed{E_T}$ from another neutrino as well as from the lost-lepton from the $W$-decay. However, the preselection criteria for $t \bar t$ leptonic is stronger compared to $t \bar t$ semileptonic case, since we demand exactly one lepton in the final state. The observable $t_{mod}$ plays important role in case of $t \bar t$ backgrounds. For $W+$jets background the preselection criteria, which demands at least one $b$-jet, itself eliminates significant fraction of events. Then $\slashed{E_T}, M_T, \Delta \phi(\vec{\ell},\vec{\slashed{E_T}})$, reduce this background even further. For $t \bar tZ$ where $t \bar t$ decays semileptonically and $Z$ decays to a pair of neutrinos and $WZ$ where $W$ decays leptonically and $Z$ decays to a pair of neutrinos, $M_T$ variable does not help, since there too, the $\slashed{E_T}$ comes from multiple sources. However, $\slashed{E_T}$ and $t_{mod}$ play important role in reducing these backgrounds. In case $WZ$ background $b$-jet identification at the preselection and top-tagging in the final state, put strong constraint. We present the number of events after applying the aforementioned cuts for 14 TeV, 33 TeV and 100 TeV in Table~\ref{tablecutflow14},~\ref{tablecutflow33} and~\ref{tablecutflow100}.

The significance~\cite{Cowan:2010js} has been calculated using the following formula.
\begin{equation}
{\cal{S}}=\sqrt{2[(S+B) Log(1+ \frac{S}{B}) -S]}
\end{equation}

 \begin{table}[!hptb]
\scriptsize{
\begin{center}
\begin{tabular}{| c | c | c | c |}
\hline
Benchmarks & $\sqrt{s}$ = 14 TeV ${\cal L}$ = 3000 $fb^{-1}$ & $\sqrt{s}$ = 33 TeV ${\cal L}$ = 1000 $fb^{-1}$ & $\sqrt{s}$ = 100 TeV ${\cal L}$ = 100 $fb^{-1}$ \\
\hline
BP1 left-handed & 2.0$\sigma$ & 24$\sigma$ & 63$\sigma$ \\
\hline
BP1 right-handed & 2.4$\sigma$ & 28$\sigma$ & 73$\sigma$ \\
\hline
BP2 left-handed & 1$\sigma$ & 12.5$\sigma$ & 35$\sigma$ \\
\hline
BP2 right-handed & 1.2$\sigma$ & 15$\sigma$ & 41$\sigma$ \\
\hline
BP3 left-handed & 0.7$\sigma$ & 4.7$\sigma$ & 9$\sigma$ \\
\hline
BP3 right-handed & 0.8$\sigma$ & 5.3$\sigma$ & 10$\sigma$ \\
\hline
BP4 left-handed & 0.4$\sigma$ & 2.5$\sigma$ & 4.8$\sigma$ \\
\hline
BP4 right-handed & 0.4$\sigma$ & 2.7$\sigma$ & 5.3$\sigma$ \\
\hline
\end{tabular}
\caption{Signal significance for various benchmarks at different values of $\sqrt{s}$ and integrated luminosity ${\cal L}$.}
\label{significance}
\end{center}
}
\end{table}

We present the signal significance for all our benchmarks at various $\sqrt{s}$ and integrated luminosities in Table~\ref{significance}. One can see that, benchmarks BP3 and BP4, which correspond to low $p_T$ top quarks, can not be probed at the 14 TeV run of LHC even with high luminosity, if one resorts to the resolved category of top-jet tagging. However, in the high energy runs of LHC it will be possible to probe these low $p_T$ top events even in the resolved category with $\sim 3-5\sigma$ significance. BP1 and BP2, on the other hand, performs slightly better($1-2\sigma$) compared to the other two benchmarks at $\sqrt{s} = 14$ TeV  and ${\cal L} = 3000 fb^{-1}$, but with increased centre of mass energy, they can be probed with large significance even with relatively lower luminosity.

\section{Polarization observables}
\label{sec5}

Having analyzed signal and background events and calculating the number of events in the signal region for $\sqrt{s}$ = 14 TeV, 33 TeV and 100 TeV, we proceed towards extracting top polarization in all these scenarios. First, we focus on an observable which can provide a good discrimination between left and right-handed hadronically decaying top quarks in the realistic LHC environment. Next, we will calculate the degree with which the aforementioned distinction can be made, in presence of the remaining background events in our chosen signal region. Interesting energy observables have been suggested and explored in an earlier work~\cite{Krohn:2009wm}. However, it has been pointed out in~\cite{Godbole:2019erb}, that those suggested observables lose their distinguishing capacity in the real environment of LHC, on applying the crucial top-tagging algorithm involving jet-substructure analysis. Certain angular observables and asymmetries, which are more suited for this purpose, were introduced and explored in~\cite{Godbole:2019erb}. In the present work, we are going to explore the reach the aforementioned observables at the future high-luminosity and high-energy LHC runs.

\subsection{Angular observable}

In order to construct this observable, one has to first identify the b-like subjet inside the tagged top, which can be done after $W$ mass reconstruction and top-tagging, out of three subjets inside the top-tagged fatjet.
Next we impose the condition, $m_{bj_1} < m_{bj_2}$, where $j_1$ and $j_2$ are the two sub-jets that constitute $W$-boson inside the fatjet. 
It has been shown in \cite{Brandenburg:2002xr,Tweedie:2014yda} that  50\% to $60 \%$ 
of the times $j_1$ acts as a proxy
for the $d$-subjet, which has the maximum spin-analyzing power, as we have discussed earlier.

The angle
between the momenta of thus identified sub-jet $j_1$ and top fatjet in the top rest frame is therefore
correlated with the polarization of the parent top quark.
The angle in discussion, $\theta$, is defined as follows.
\begin{eqnarray}
\cos \theta & \equiv &
\frac{ {{\vec{ t_j} }} \cdot {{\vec{j}_1}} }
{ \left| {{\vec{t_j}}} \right|\left| {{\vec{j}_1}} \right|},
\label{eq:angle}
\end{eqnarray}
where ${\vec{t}_j}$ is the momentum of the reconstructed 
top fatjet in the lab frame, and 
${\vec{j}_1}$ is the momentum of the sub-jet $j_1$ in top jet rest frame.

\begin{figure}[!hptb]
	\centering
\includegraphics[width=7cm,height=6cm]{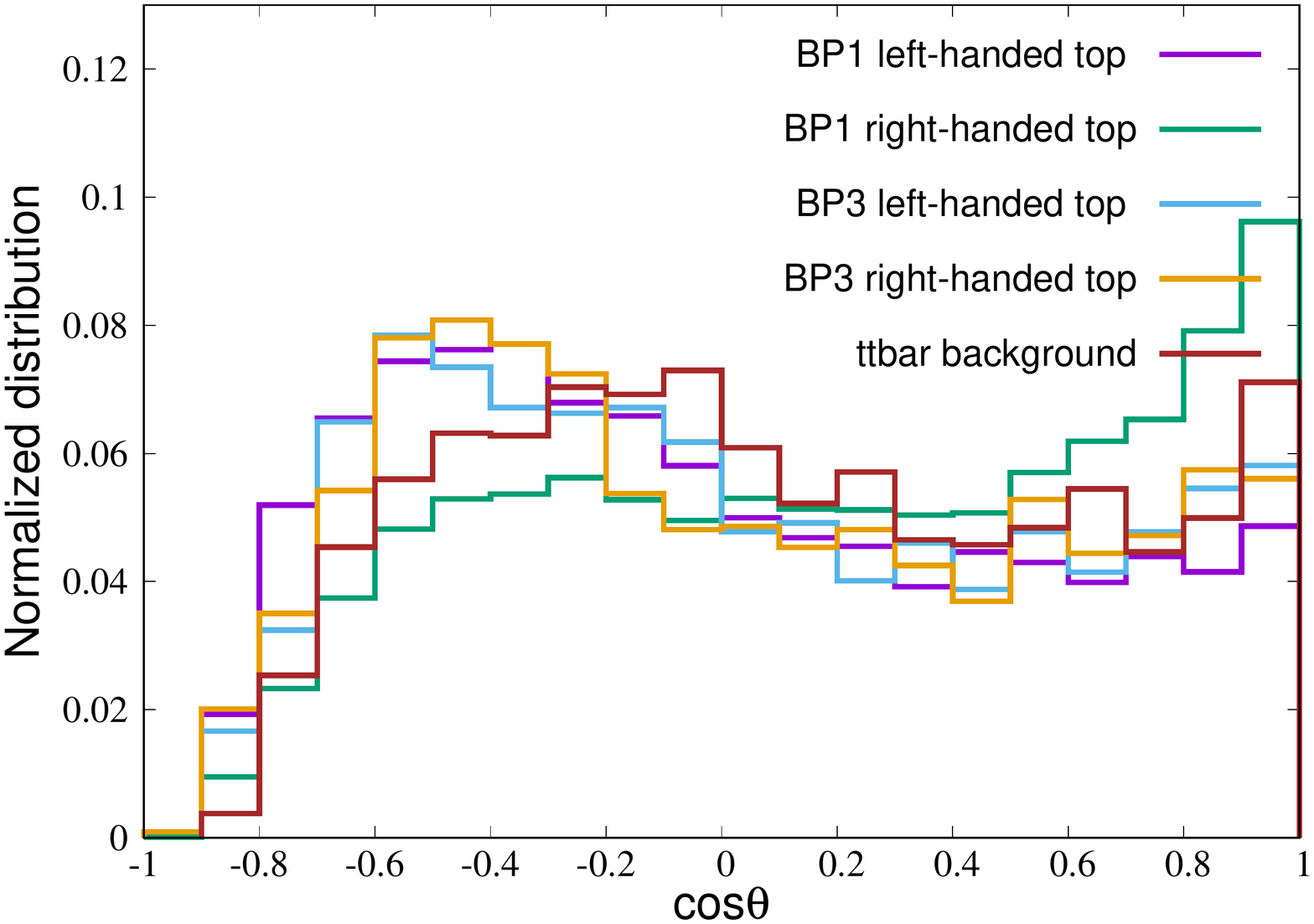}
\includegraphics[width=7cm,height=6cm]{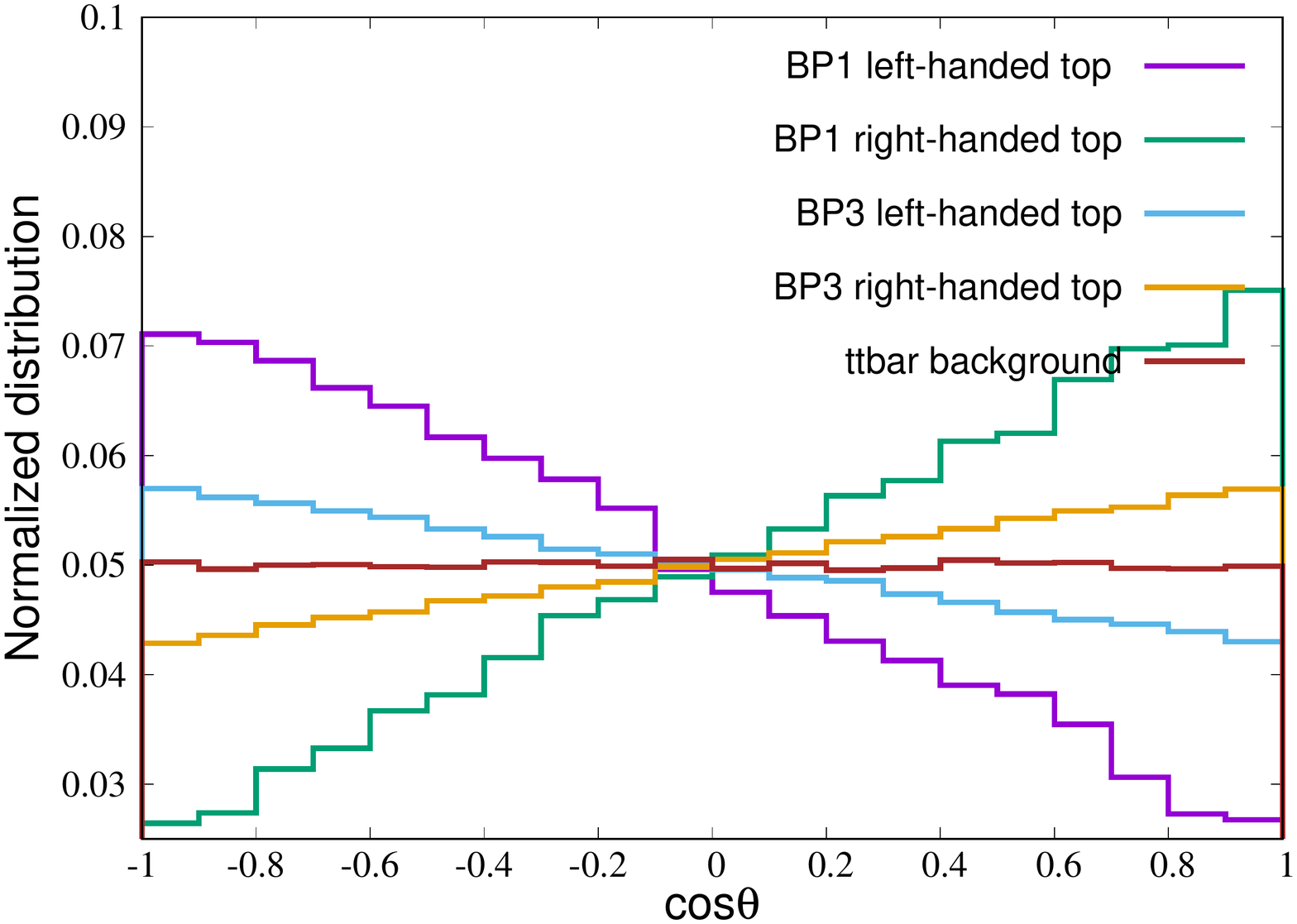}
	\caption{Normalized distribution of $\cos \theta$(left) for signal and $t \bar t$ background. (right) Normalized $\cos\theta$ distribution at the parton level. }
	\label{costheta}
\end{figure}

In Figure~\ref{costheta}(left), we show normalized angular distribution for left- and right-handed top from the signal as well as the unpolarized top from the dominant $t \bar t$ background. One can see, in case of left-polarized top the distribution peaks in the backward region and in case of right-handed top the distribution peaks in the forward region, a behavior expected from Equation~\ref{pol}. In case of unpolarized top the distribution is slightly more uniform over the entire range of $\theta$. We show the corresponding parton-level distributions in Figure~\ref{costheta}(right). One can clearly see, that for BP3(similarly in BP4), the top polarization takes smaller value compared to BP1(also BP2). The smaller boost in case of BP3 and BP4 is in fact responsible for a decrease in top polarization in the helicity basis. We have checked that in the helicity basis, at the parton level, top quark acquires 50\% polarization in BP1 and BP2, whereas in case of BP3 and BP4 it decreases to 15\%.   

\subsection{Asymmetry}

Guided by the angular distributions of the left-handed, right-handed and unpolarized top, one can construct asymmetries, which are sensitive to the couplings of the top quark. The forward backward asymmetry in this case will be defined as

\begin{equation}
A_{\theta} = \frac{N_{\cos \theta > 0} - N_{\cos \theta < 0}}{N_{\cos \theta > 0} + N_{\cos \theta < 0}}
\end{equation}

One can easily check from Equation~\ref{pol} that the asymmetry $A_{\theta}$ is actually proportional to the polarization ${\cal P}$ of the top quark. So the asymmetry can be used as a direct measure of the polarization. In~\cite{Godbole:2019erb}, variation of the asymmetry with different mixing parameters in the stop and neutralino sector has been explored in detail. However, in the presence of background events, the situation is different. It is worthwhile studying the distribution and corresponding asymmetries in the presence of remaining background events after all the kinematic cuts are applied. In Table~\ref{asymmetry}, we quote the asymmetry values for our signal benchmarks at various centre of energies and integrated luminosities.

\begin{table}[!hptb]
\scriptsize{
\begin{center}
\begin{tabular}{| c | c | c | c |}
\hline
Benchmarks & $\sqrt{s}$ = 14 TeV ${\cal L}$ = 3000 $fb^{-1}$ & $\sqrt{s}$ = 33 TeV ${\cal L}$ = 1000 $fb^{-1}$ & $\sqrt{s}$ = 100 TeV ${\cal L}$ = 100 $fb^{-1}$ \\
\hline
BP1 left-handed & 0.109 & -0.057 & -0.118 \\
\hline
BP1 right-handed & 0.131 & 0.122 & 0.176 \\
\hline
BP2 left-handed & 0.119 & -0.028 & -0.090 \\
\hline
BP2 right-handed & 0.124 & 0.111 & 0.149 \\
\hline
BP3 left-handed & 0.113 & 0.016 & -0.013 \\
\hline
BP3 right-handed & 0.126 & 0.087 & 0.118 \\
\hline
BP4 left-handed & 0.118 & 0.032 & 0.015 \\
\hline
BP4 right-handed & 0.128 & 0.067 & 0.087 \\
\hline
\end{tabular}
\caption{Forward-backward asymmetry for various benchmarks at different values of $\sqrt{s}$ and ${\cal L}$.}
\label{asymmetry}
\end{center}
}
\end{table}

At the parton-level, the left(right)-polarized top should give rise to negative(positive) values of asymmetry, which can be calculated from Equation~\ref{pol}. Even after showering, hadronization, fatjet formation and consequent top-tagging, left(right)-handed top from stop decay will give rise to negative(positive) asymmetry, albeit differing in magnitude from its parton-level values. However, the situation will be quite different in the presence of background events. One can see that the major $t \bar t$ background, being asymmetric (see Figure~\ref{costheta}) with respect to $\theta = 0$, affects the signal events in the forward($\cos\theta > 0)$ and backward($\cos\theta < 0$) region differently. The background adds more events to the backward region compared to that in the forward region. Therefore, the magnitude of asymmetry decreases compared to the signal-only scenario. For example, in the left-handed cases, especially at $\sqrt{s}$ = 14 TeV, where the background events are dominant, the asymmetry ends up taking positive value due to the aforementioned reason. One should note that, with increasing $\sqrt{s}$ the effect of background becomes less prominent and asymmetries tend to behave like the signal-only case, which is most evident from the 100 TeV results. 

\section{Calculation of $P$-value and significance level}
\label{sec6}

In this section we will perform a goodness-of-it test on the $\cos \theta$ distributions. For the test, one has to first start with a histogram of $\cos\theta$ with $N$-bins. In order to calculate the degree with which left- and right-handed top quarks in various benchmarks can be distinguished, we first identify the expected and observed number of events in each bins. We consider one handedness as hypothesis(expected) and the opposite handedness as truth(observed) on a case-by-cases basis. Supposing $n_i$ is the observed number of events and $\nu_i$ is the expected number of events in the $i$-th bin, we would like to construct a statistic which will reflect the level of (dis)agreement between the truth and hypothesis. For this purpose, we use the test statistic based on Pearson's $\chi^2$-statistic, which is defined as follows.

\begin{equation}
\chi^2 = \sum_{i=1}^N \frac{(n_i-\nu_i)^2}{\nu_i}
\label{chisquare}
\end{equation} 

\noindent
If the $n_i$ events in each bins are Poisson-distributed with mean $\nu_i$ and the number in each bins are not too small ($n_i > 5)$, one can show that the quantity in Equation~\ref{chisquare} follows a $\chi^2$ distribution with $N$ degrees of freedom~\cite{Cowan:1998ji}. This statement holds regardless of the distribution. 

Since the Poisson-variable with mean $\nu_i$ will have standard deviation $\sqrt{\nu_i}$, the quantity in Equation \ref{chisquare} is nothing but the sum of deviations from the expected values in each bin in units of standard deviation of that particular bin. Evidently, the larger the $\chi^2$, higher is the disagreement between the hypothesis and truth. In order to quantify this disagreement one has to calculate the $P$-value or significance level which is defined as the integral of the $\chi^2$ distribution from the specific $\chi^2$ obtained from Equation~\ref{chisquare} to infinity. 

\begin{equation}
P = \int_{\chi^2}^\infty f(z, N) dz
\label{pvalue}
\end{equation}

\noindent
The $\chi^2$ distribution function $f(z,N)$ for $N$ degrees of freedom and a continuous variable $z (0 \leq z < \infty$) is given as follows:

\begin{equation}
f(z,N) = \frac{1}{2^{N/2}\Gamma(N/2)}z^{N/2 - 1}e^{-z/2} 
\end{equation}

\noindent
The definition uses Gamma function where $\Gamma(x) = \int_0^\infty e^{-t}t^{x-1} dt$.

\begin{figure}[!hptb]
	\centering
   \includegraphics[width=7cm,height=6cm]{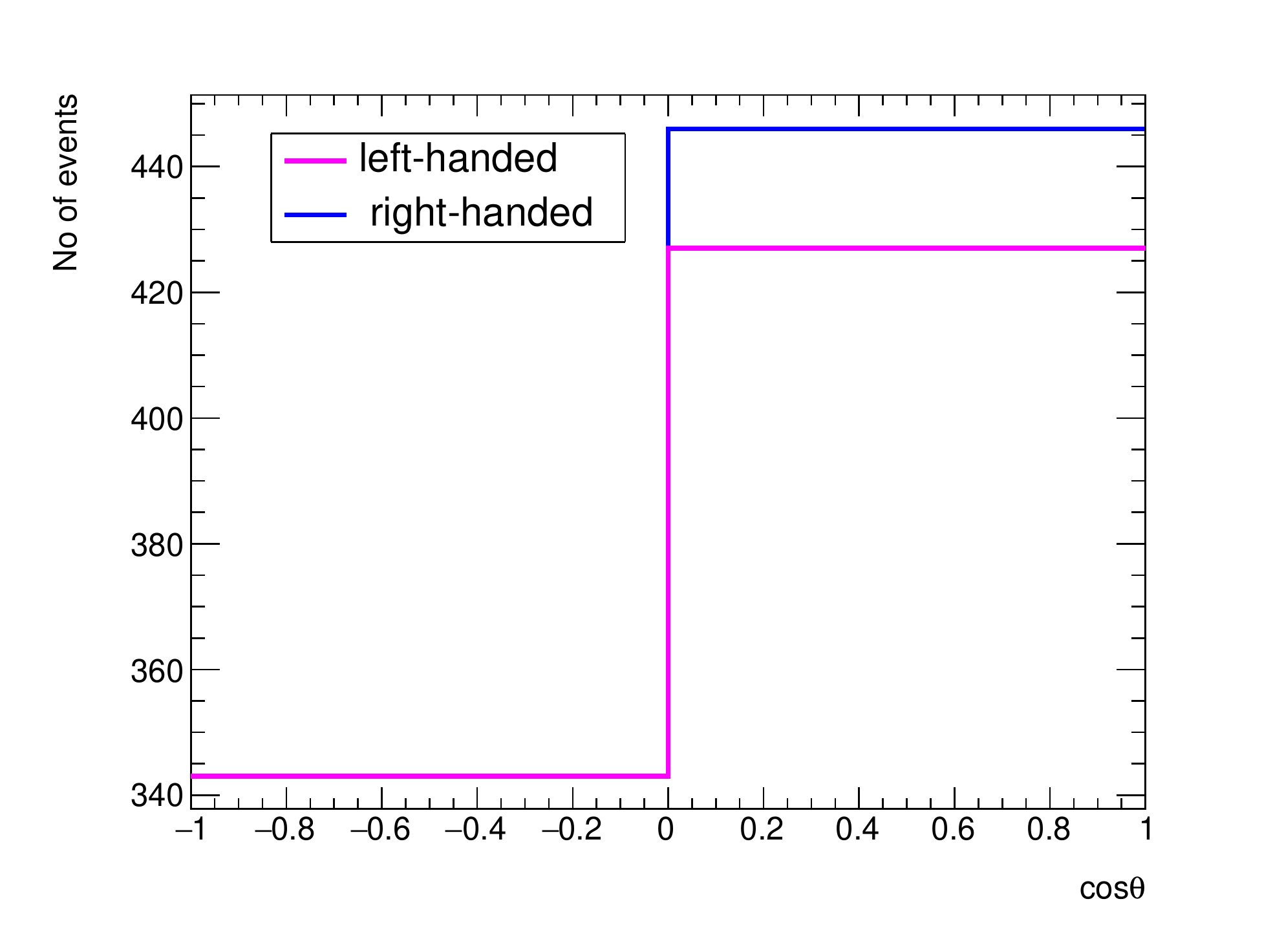}
\includegraphics[width=7cm,height=6cm]{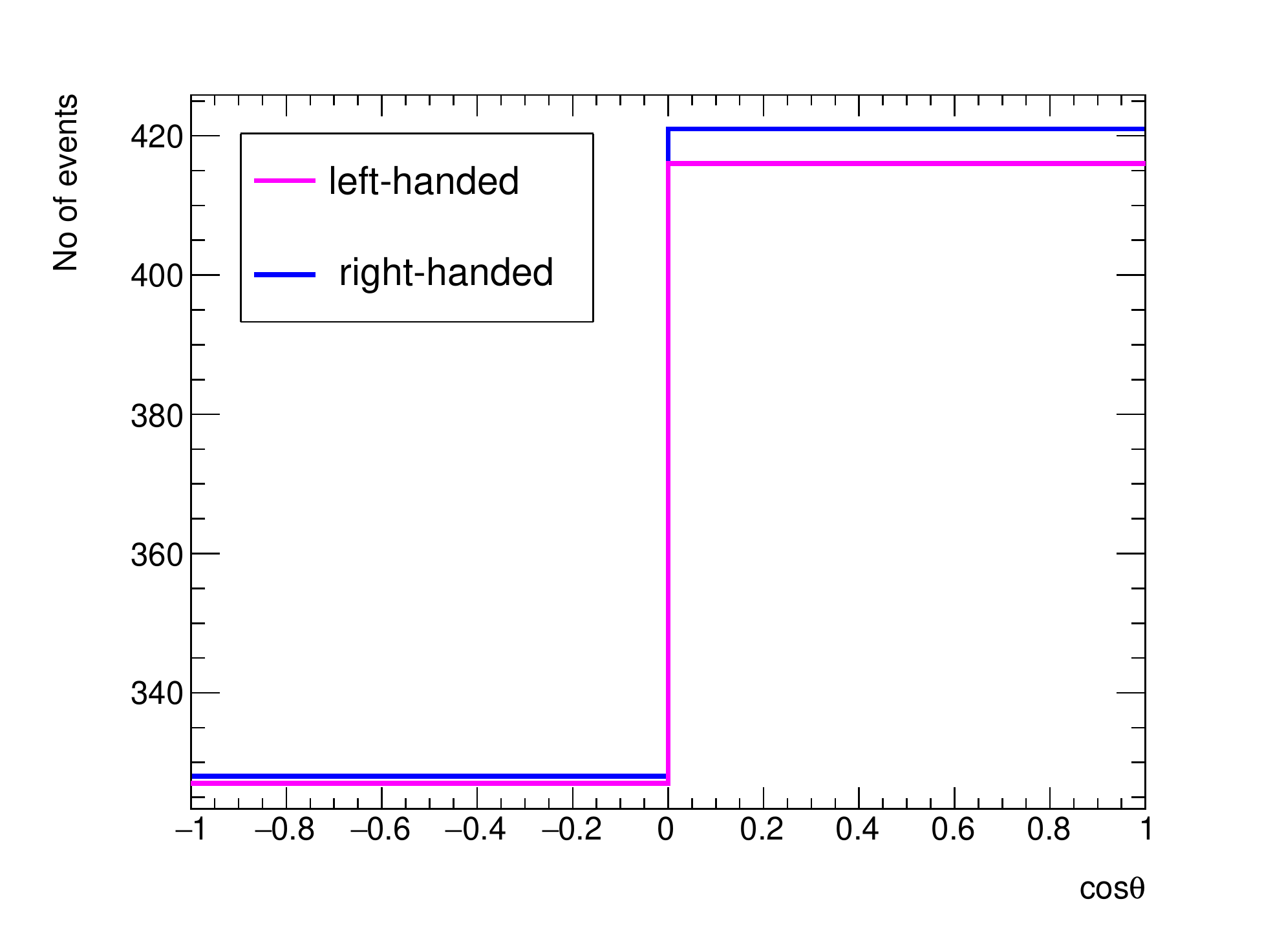} \\
  \includegraphics[width=7cm,height=6cm]{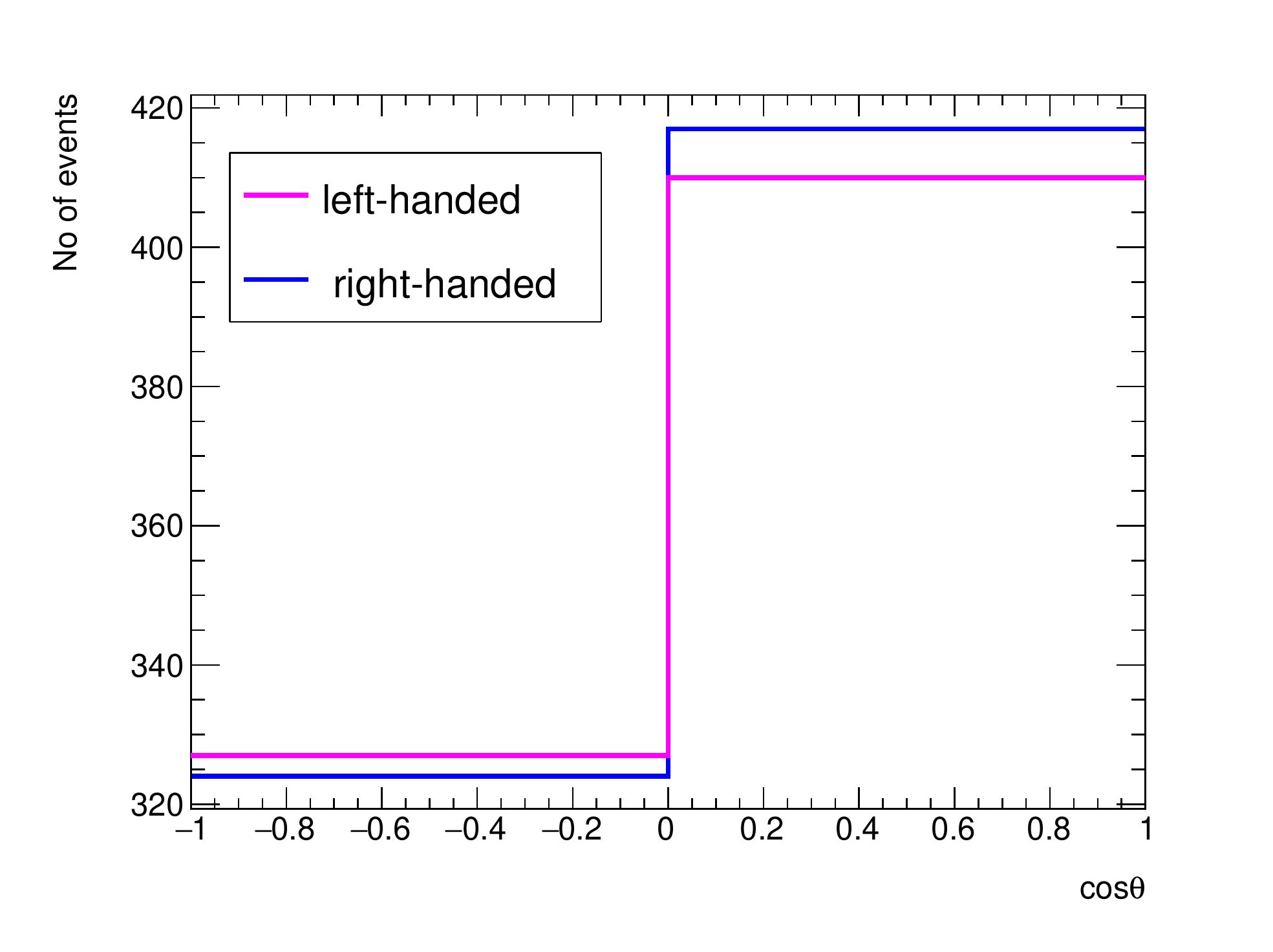}
\includegraphics[width=7cm,height=6cm]{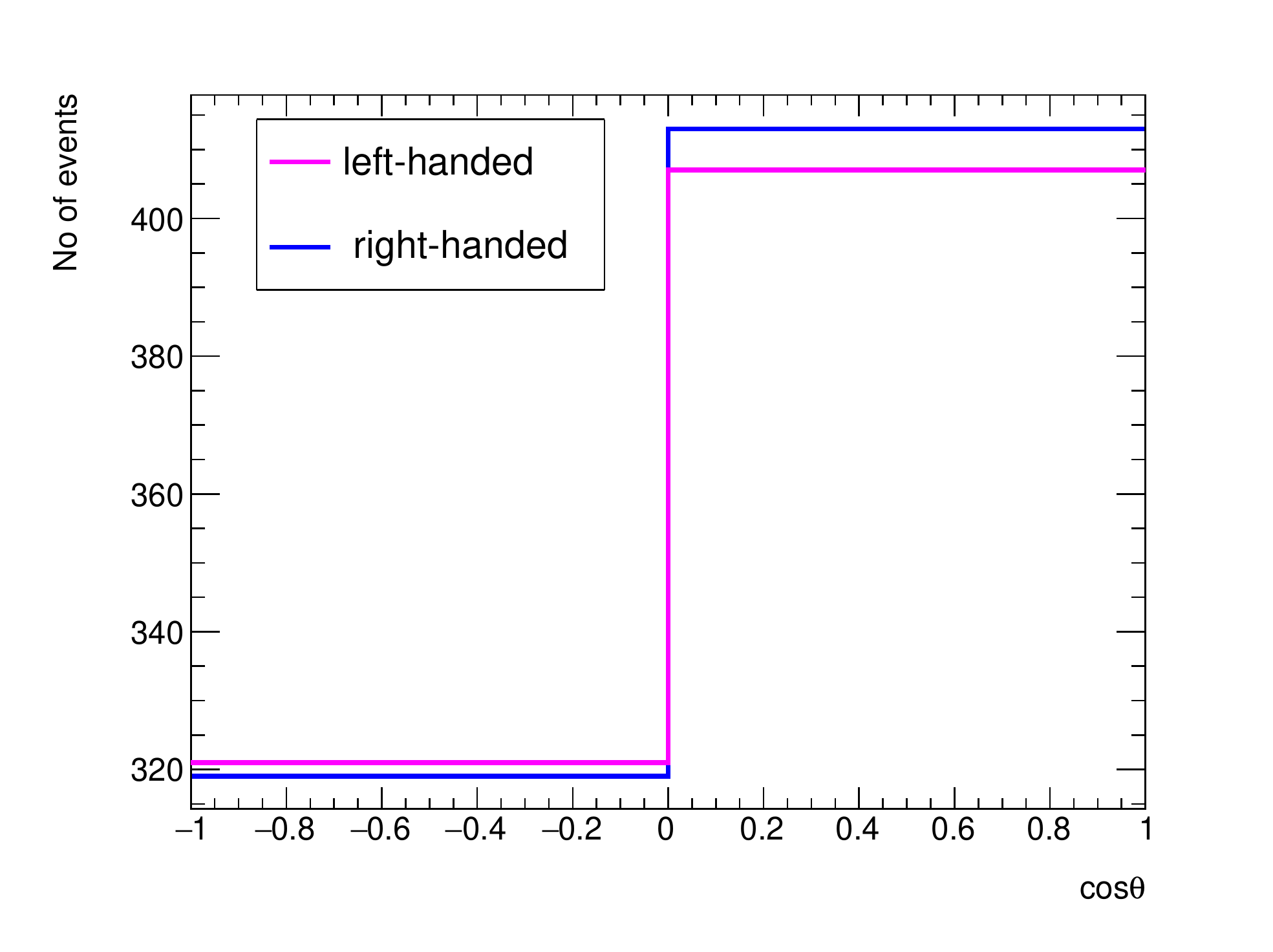} \\
	\caption{Sum of signal + background events after the application of all the  aforementioned cuts for both left- and right-handed top from stop decay at $\sqrt{s} = 14$TeV and 3000$fb^{-1}$.}
	\label{stat_14tev}
\end{figure}

 \begin{table}[!hptb]
\begin{center}
\begin{tabular}{| c | c | c | }
\hline
 & Hypothesis left-handed, Truth right-handed & Hypothesis right-handed, Truth left-handed \\
\hline
BP1 & $P$-value = 0.65($0.47\sigma$) & $P$-value = 0.67($0.45\sigma$)\\
\hline
BP2  &  $P$-value = 0.96($0.05\sigma$) & $P$-value = 0.96($0.05\sigma$)\\
\hline
BP3  &  $P$-value = 0.92($0.12\sigma$) & $P$-value = 0.92(0$.12\sigma$)\\
\hline
BP4 &  $P$-value = 0.95($0.07\sigma$) & $P$-value = 0.95($0.07\sigma$)\\
\hline
\end{tabular}
\caption{Signal significance for various benchmarks at different values of $\sqrt{s} = 14$TeV and integrated luminosity ${\cal L}=3000 fb^{-1}$.}
\label{14tev}
\end{center}
\end{table}

We first show the $\cos\theta$ distribution for all the benchmarks at various centre-of-mass energies and integrated luminosities, on which we will perform the goodness-of-fit test following the method described above. We consider number of bins $N=2$ to minimize fluctuation in all the bins. In Figure~\ref{stat_14tev}, we show the distribution of signal + background for all the signal benchmarks at $\sqrt{s} = 14$ TeV and ${\cal L} = 3000 fb^{-1}$ after using all the analysis cuts. It is evident from the figure that the left- and right-handed top quarks in this case do not show much difference from each other. The reason behind that is at 14 TeV the background distribution dominates over the signal and the resulting distribution for both left- and right-handed case essentially corresponds to that of the background, a behavior which was apparent in the calculation of asymmetry as well in the preceding section. One should note that the benchmarks BP3 and BP4 suffers from two issues which results in reduced distinguishability for these benchmarks. BP3 and BP4 corresponds to top quarks with lower boost and therefore the polarization of top in the helicity basis is somewhat smaller compared to BP1 and BP2, a feature we have discussed earlier. On the other hand, low $p_T$ of top quark also results in lowered top-tagging efficiency, leading to reduced signal yield (see Table~\ref{tablecutflow14}). We present the corresponding $P$-value and significance level in Table~\ref{14tev}, which clearly shows that 14 TeV high luminosity LHC will not be able to distinguish between the left- and right-handed tops arising from different types of interactions between stop-top-neutralino.

\begin{figure}[!hptb]
	\centering
   \includegraphics[width=7cm,height=6cm]{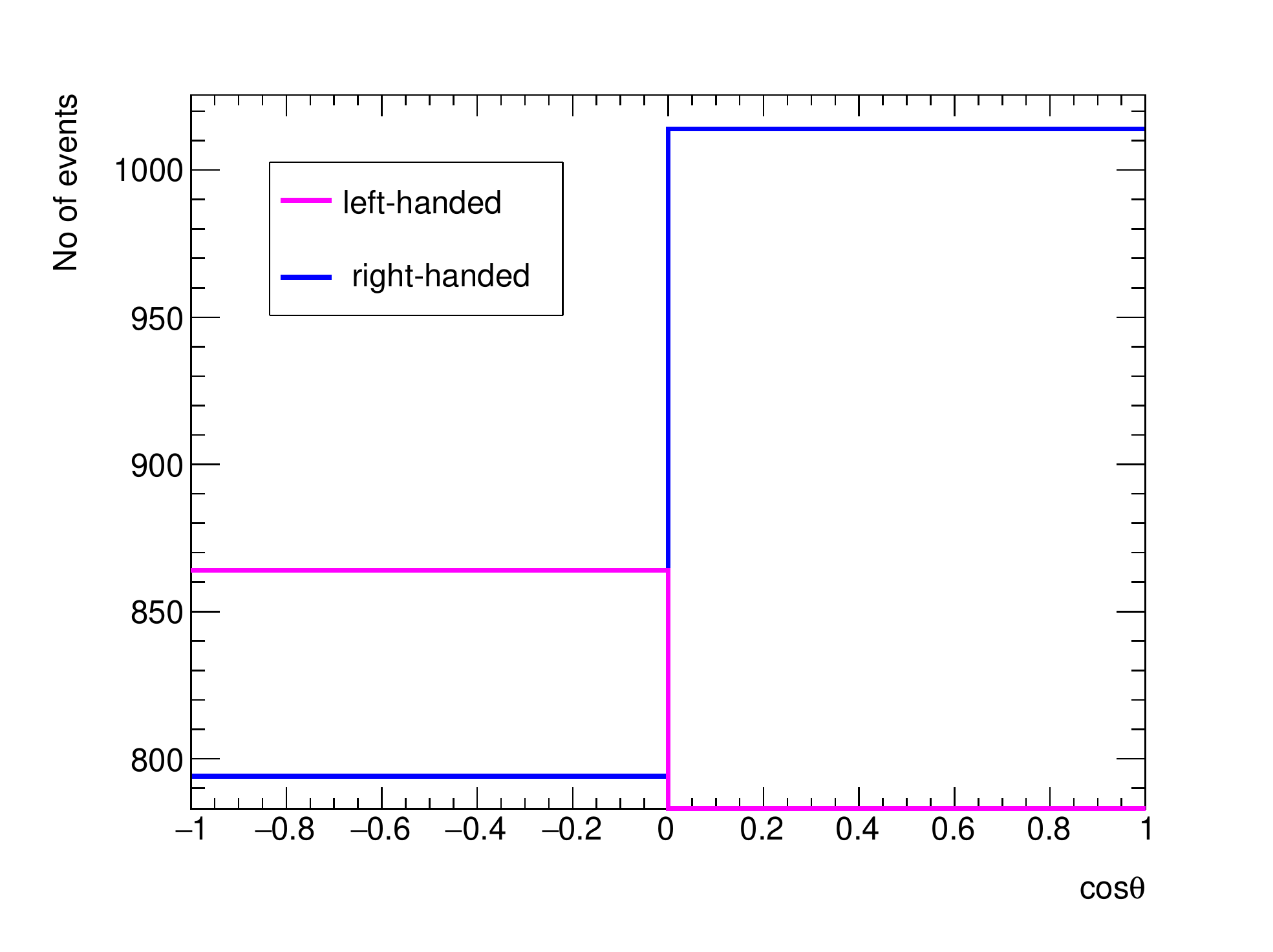}
\includegraphics[width=7cm,height=6cm]{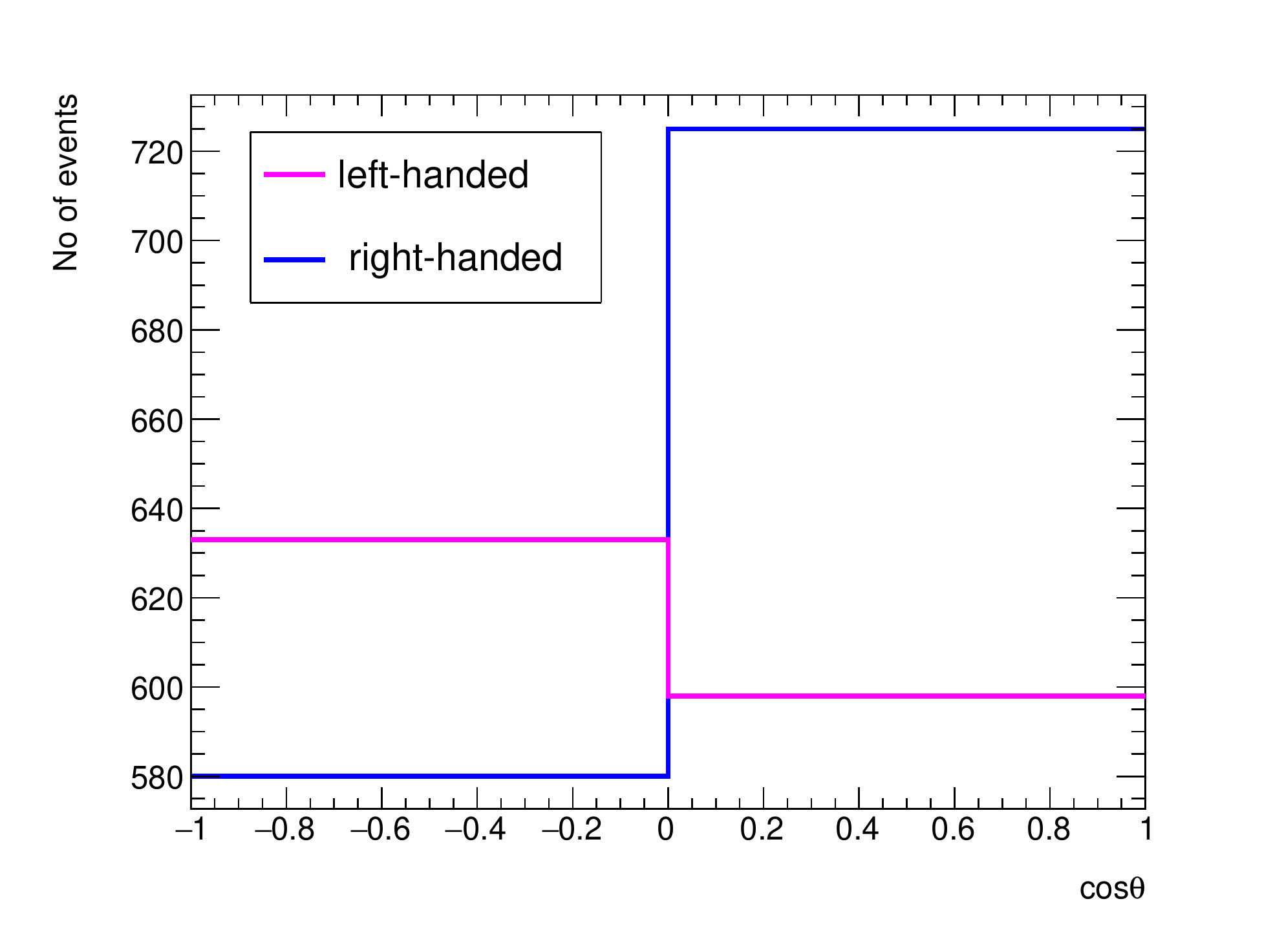} \\
  \includegraphics[width=7cm,height=6cm]{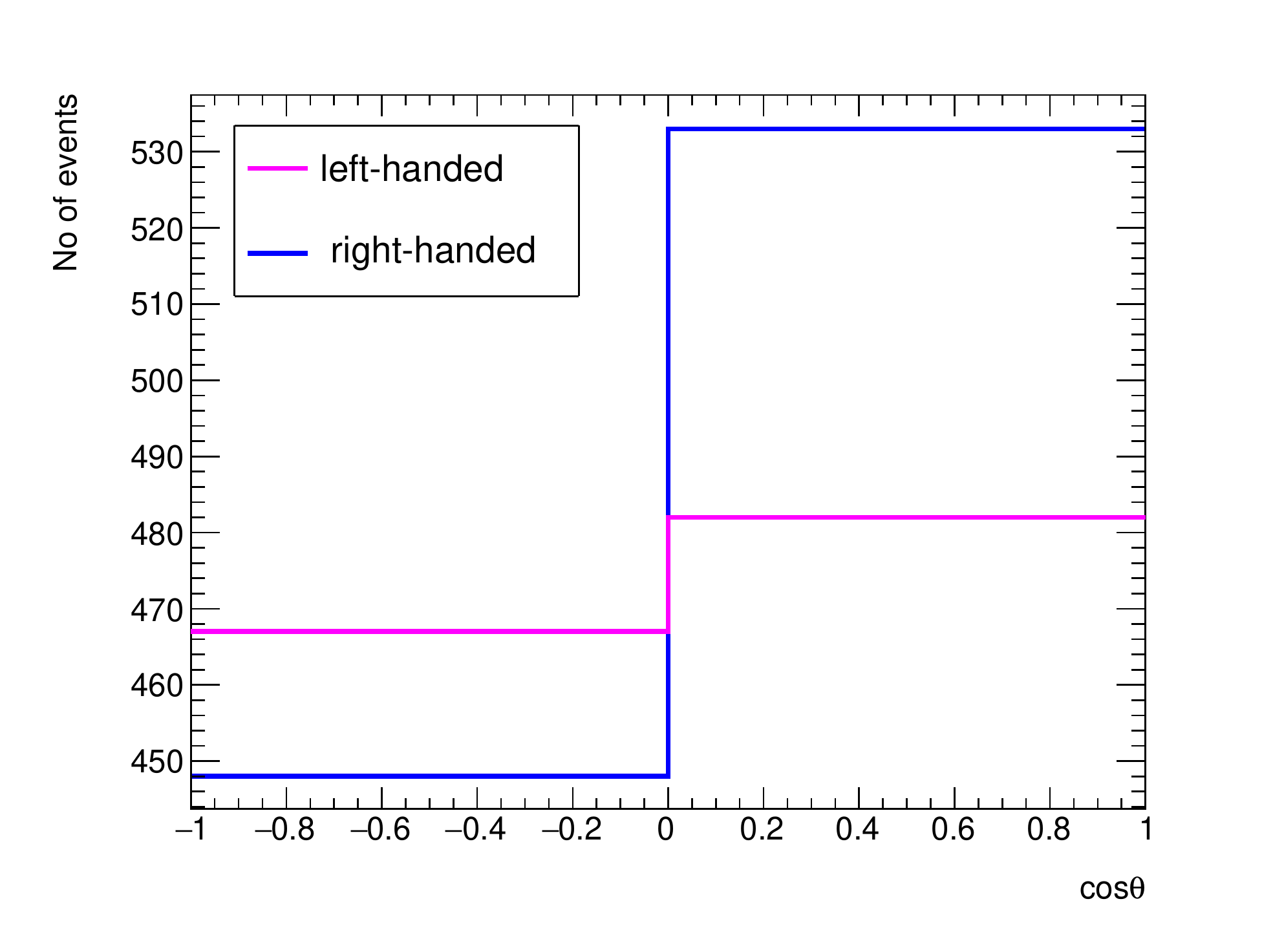}
\includegraphics[width=7cm,height=6cm]{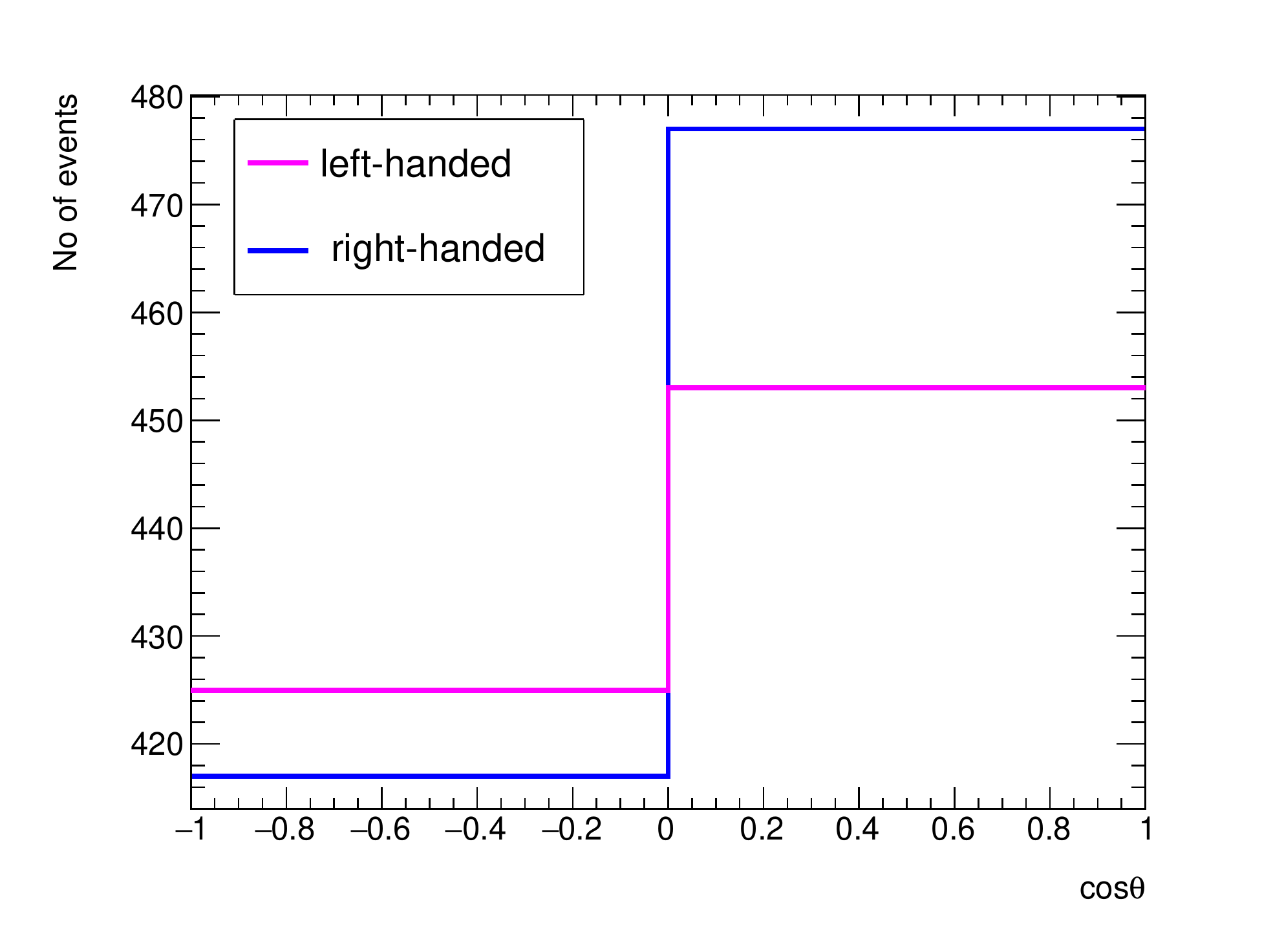} \\
	\caption{Sum of signal + background events after the application of all the  aforementioned cuts for both left- and right-handed top from stop decay at $\sqrt{s} = 33$TeV and 1000$fb^{-1}$.}
	\label{stat_33tev}
\end{figure}

 \begin{table}[!hptb]
\begin{center}
\begin{tabular}{| c | c | c | }
\hline
 & Hypothesis left-handed, Truth right-handed & Hypothesis right-handed, Truth left-handed \\
\hline
BP1 & $P$-value = 1.0$\times 10^{-18}$($9.0\sigma$) & $P$-value = 1.1$\times 10^{-14}$(8.0$\sigma$)\\
\hline
BP2  &  $P$-value = 1.5$\times 10^{-7}$($5.3\sigma$) & $P$-value = 1.3$\times 10^{-6}$ $(4.9\sigma$)\\
\hline
BP3  &  $P$-value = 0.045(2.0$\sigma$) & $P$-value = 0.058(1.9$\sigma$)\\
\hline
BP4 &  $P$-value = 0.4($1.0\sigma$) & $P$-value = 0.45($0.8\sigma$)\\
\hline
\end{tabular}
\caption{Signal significance for various benchmarks at different values of $\sqrt{s} = 33$TeV and integrated luminosity ${\cal L}=1000 fb^{-1}$.}
\label{33tev}
\end{center}
\end{table}

In Figure~\ref{stat_33tev}, we show the signal+background distribution for all the benchmarks at $\sqrt{s} = 33$ TeV and ${\cal L} = 1000fb^{-1}$. It is evident from the distributions, that the separation between left- and right-handed top quarks has increased significantly from the 14 TeV case. The improved signal rate over and above the background events is responsible for the improvement. We quote the relevant $P$-values and significance levels in Table~\ref{33tev}. One can see that BP1 and BP2 can reach an impressive  distinguishability of $\gsim 5\sigma$. However, BP3 and BP4 still underperforms in this case, although their reach has improved significantly from the previous case.

\begin{figure}[!hptb]
	\centering
   \includegraphics[width=7cm,height=6cm]{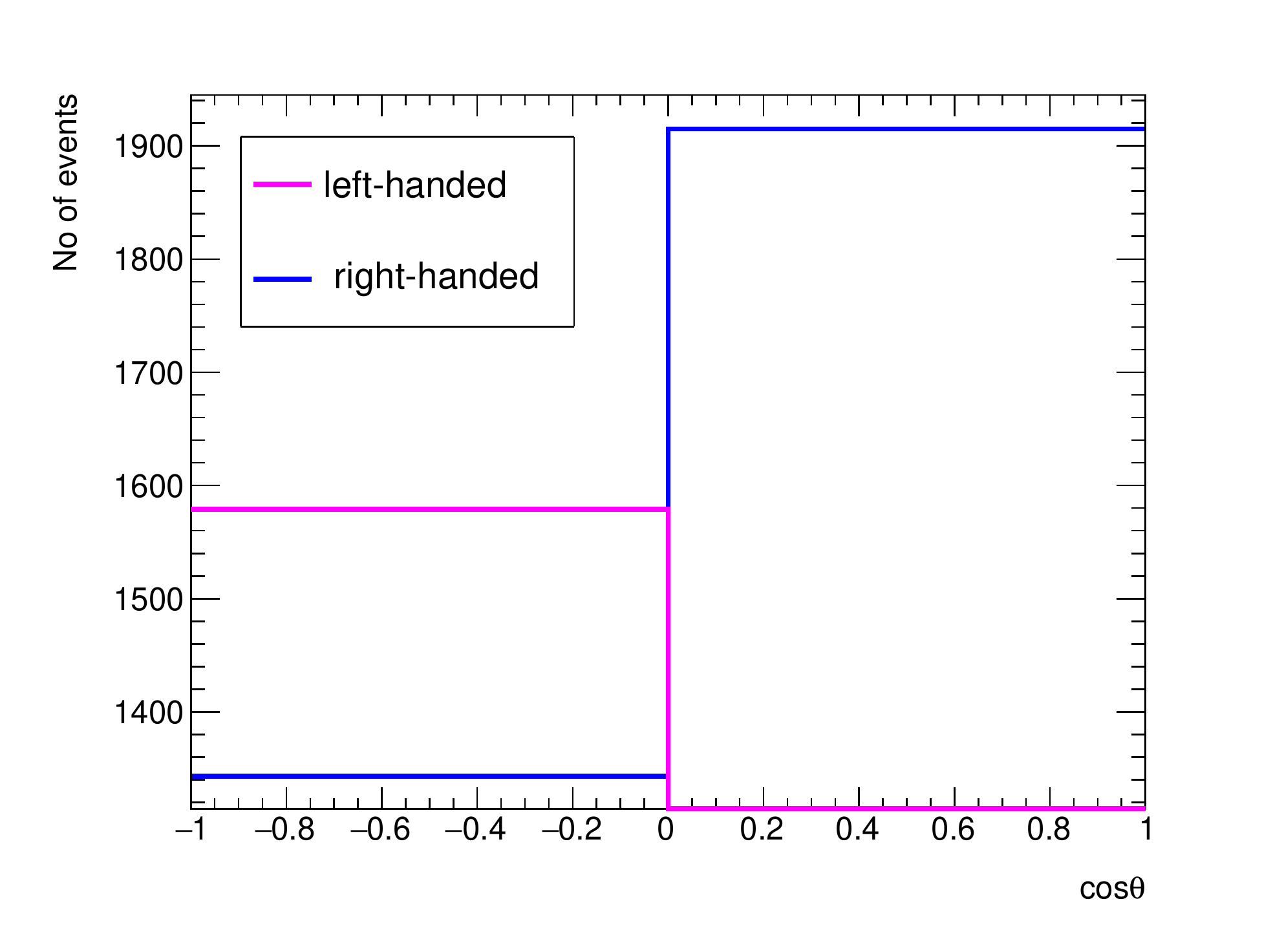}
\includegraphics[width=7cm,height=6cm]{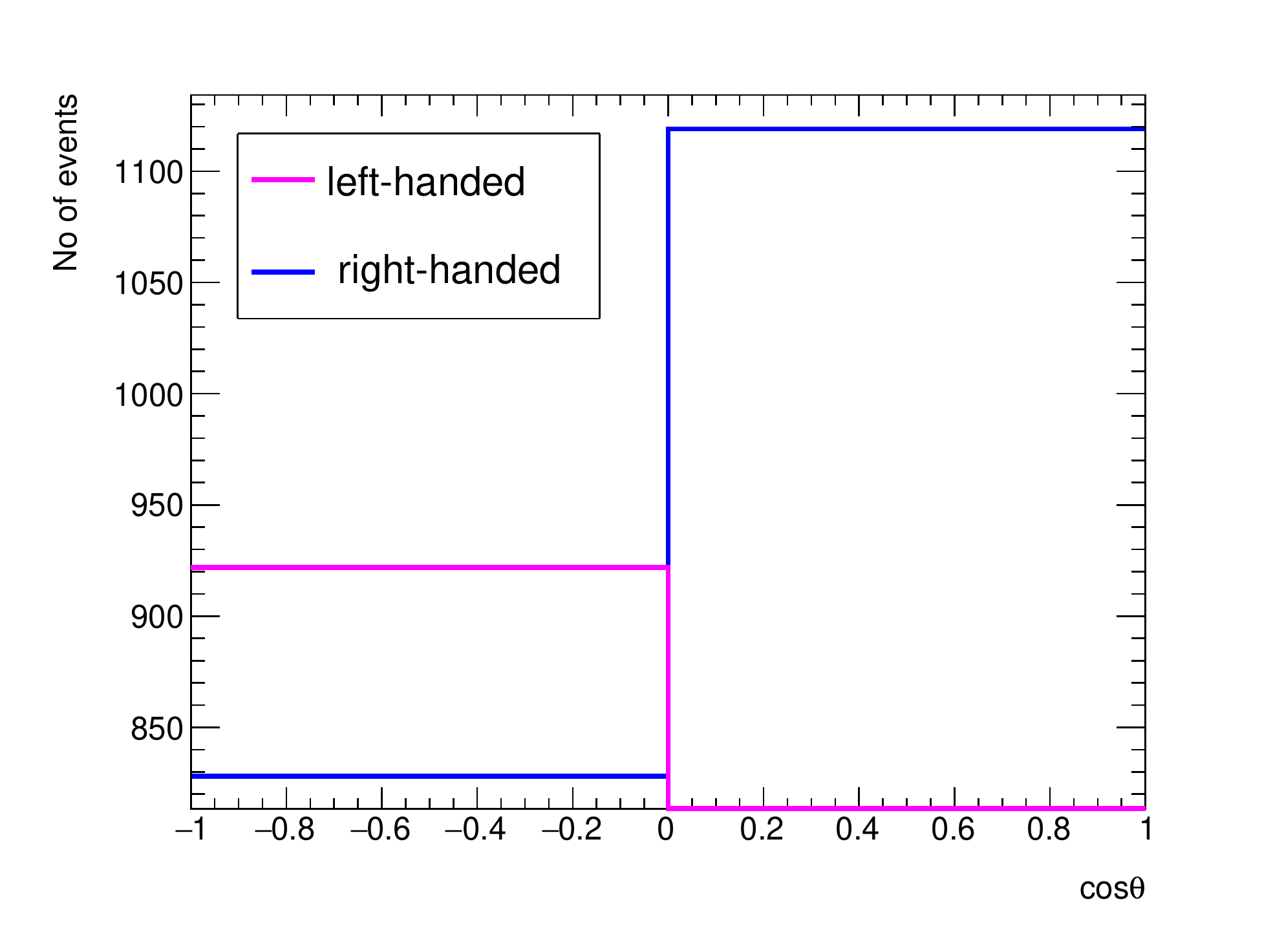} \\
  \includegraphics[width=7cm,height=6cm]{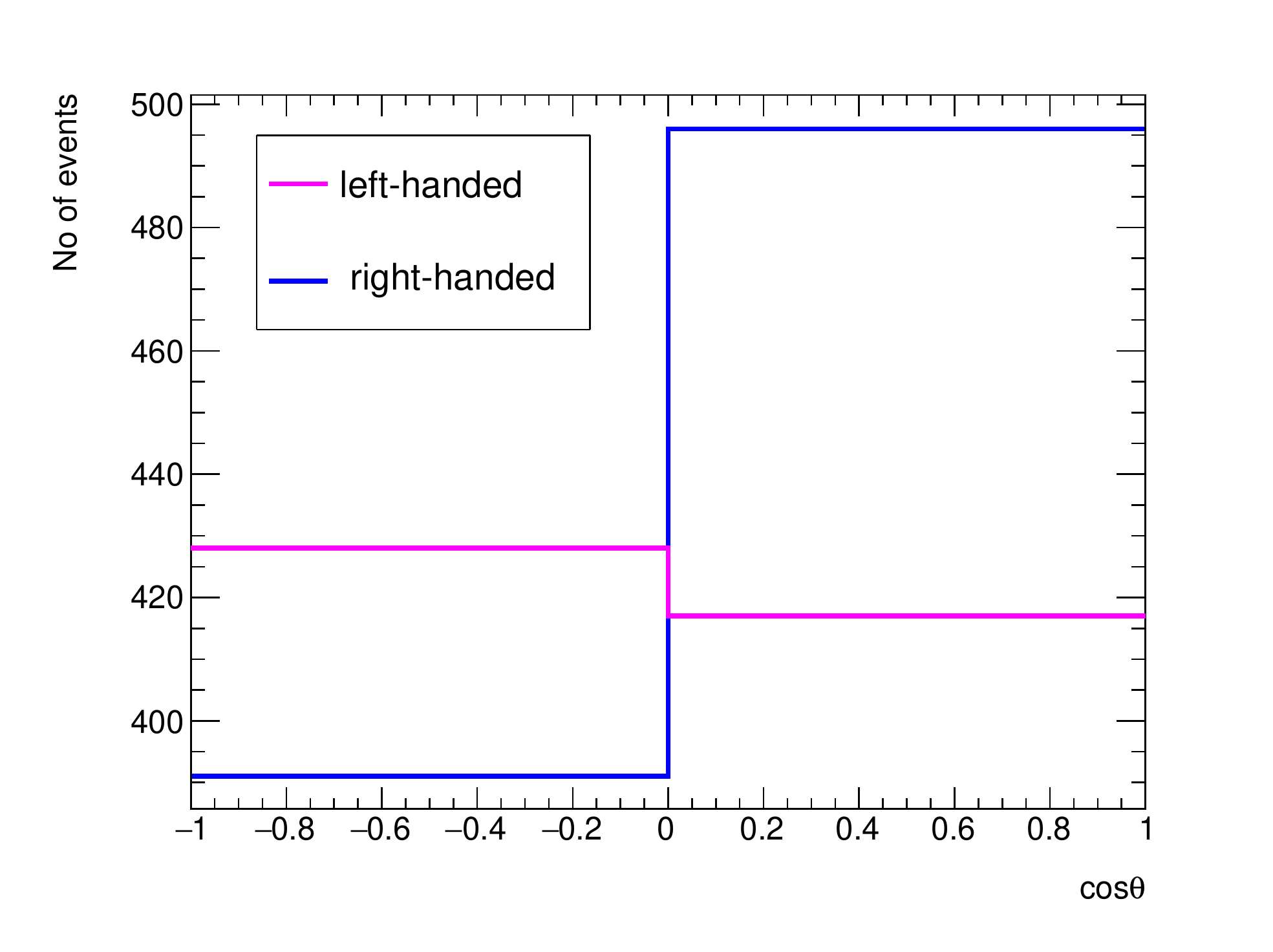}
\includegraphics[width=7cm,height=6cm]{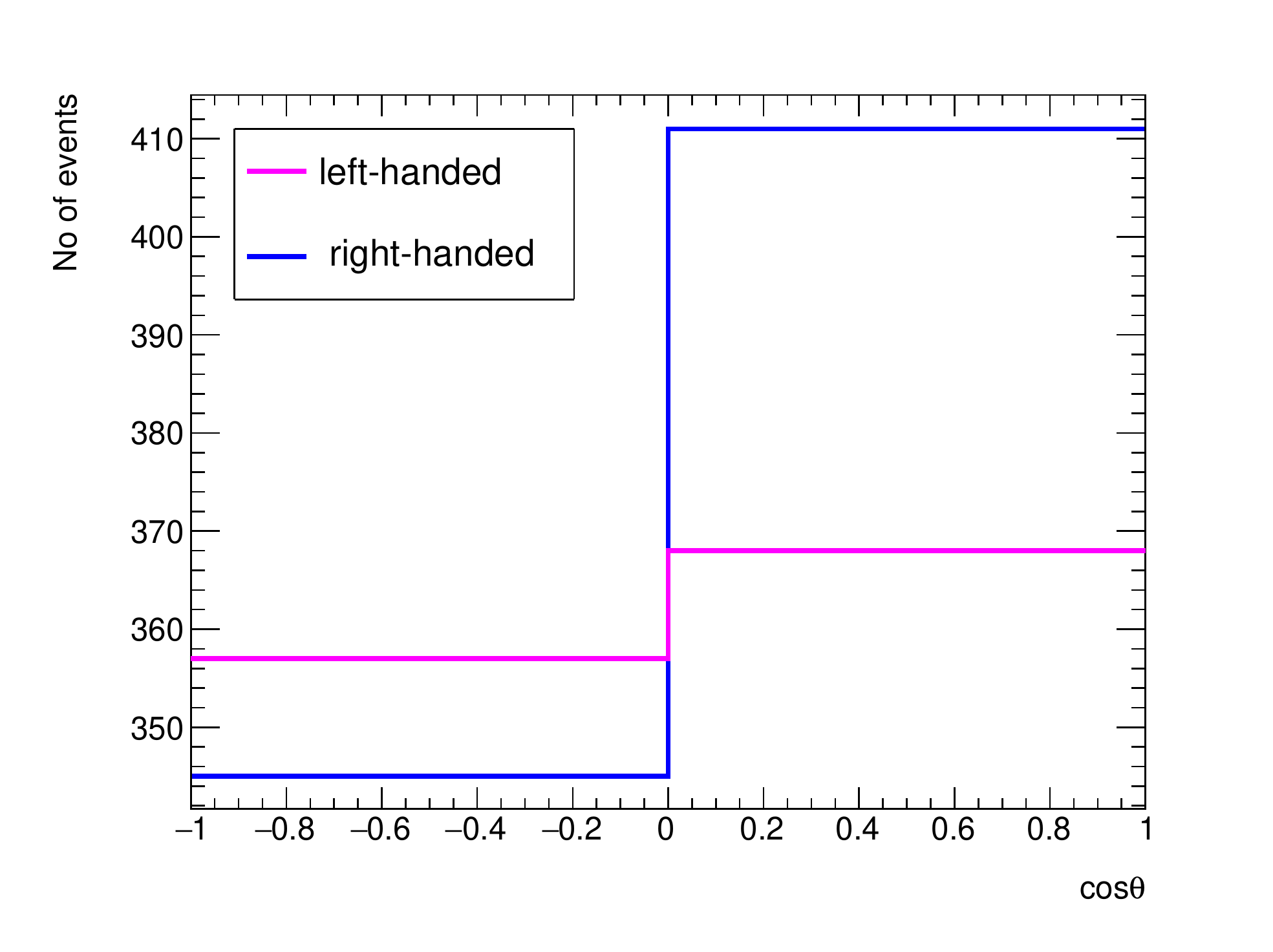} \\
	\caption{Sum of signal + background events after the application of all the  aforementioned cuts for both left- and right-handed top from stop decay at $\sqrt{s} = 100$TeV and 100$fb^{-1}$.}
	\label{stat_100tev}
\end{figure}

 \begin{table}[!hptb]
\begin{center}
\begin{tabular}{| c | c | c | }
\hline
 & Hypothesis left-handed, Truth right-handed & Hypothesis right-handed, Truth left-handed \\
\hline
BP1 & $P$-value = 1.0$\times 10^{-85}$($20\sigma$) & $P$-value = 1.0$\times 10^{-59}$($17\sigma$)\\
\hline
BP2  &  $P$-value = 1.0$\times 10^{-36}$($13\sigma$) & $P$-value = 1.0$\times 10^{-26}$($11\sigma$)\\
\hline
BP3  &  $P$-value = 1.1$\times 10^{-4}(3.9\sigma$) & $P$-value = 3.2$\times 10^{-4}(3.6\sigma$) \\
\hline
BP4 &  $P$-value = 0.067($2\sigma$) & $P$-value = 0.086($1.8\sigma$)\\
\hline
\end{tabular}
\caption{Signal significance for various benchmarks at different values of $\sqrt{s} = 100$TeV and integrated luminosity ${\cal L}=100 fb^{-1}$.}
\label{100tev}
\end{center}
\end{table}

\noindent
In Figure~\ref{stat_100tev}, we plot $\cos\theta$ distribution for all the benchmarks at 100 TeV centre-of-mass energy and 100$fb^{-1}$ integrated luminosity. We can see that in this case, the left- and right-handed distributions show clear distinction even in the presence of background events. The $P$-values corresponding to BP1 and BP2 are extremely small and the corresponding significances are large, which means left- and right-handed tops can be distinguished easily at the 100 TeV LHC, even at the early stage of run (100$fb^{-1}$).

\section{Summary and Conclusion}
\label{sec7}

We have performed a detailed collider analysis to extract polarization of boosted top quark from stop decay at the high luminosity and high energy LHC. We have discussed the phenomenology in the stop-neutralino sector of MSSM as well as resulting polarization of top quark in various regions of the parameter space. We explored the polarization measurement of the hadronically decaying top quark in the helicity basis, utilizing jet-substructure analysis. The analysis has been done in the final state consisting of one lepton + at least one $b$-jet + at least one top-tagged fatjet + $\slashed{E_T}$. We take into account the background processes which will contribute to the same final state and employ suitable kinematical cuts on kinematic observables to achieve a desired signal-background separation. Next we use angular observables and asymmetries to distinguish between left- and right-handed polarization in the signal region. Next we quantified the distinction between left- and right-handed top quarks pertaining to our chosen benchmarks, which can be achieved at the possible upgradation of LHC in the luminosity and energy frontier. This distinctions for all the cases are presented in terms of $P$-values and corresponding significance.

The decay products of boosted top quark are supposed to be confined within a cone forming a fatjet. Our analysis uses top-tagging in the boosted regime, which involves the analysis in the merged category using jet-substructure. However, the applicability of the top-tagging in the merged category, depends on the boost of the top quark. Therefore, it is of course a valid question to ask, whether our benchmarks satisfy this criteria ie. if the top quark carries significant $p_T$. Here we have focussed on such benchmarks which are detectable at the early stages of upgraded high luminosity and high energy LHC runs and explored the reach of boosted analysis in terms of signal significance as well as extracting polarization information of the top in the semileptonic final state. 

This analysis shows the prospect of proposed high energy or high luminosity runs in boosted top polarization measurement from stop decay. However, we emphasize here that this analysis is not only valid for stop decay in MSSM. Rather, similar strategy can be used to probe top quark polarization in many other new physics models, which give rise to boosted polarized top quarks. We also mention that we have considered only the semileptonic decay mode of top quark for the reasons discussed earlier. However, with increasing boost of the top quark produced from the decay of heavier states, the identification and isolation of the lepton in the final state will become more and more difficult. Under such conditions, the fully hadronic final state will probably be better-suited for discovery of the final state of interest. A full background analysis should be performed for that particular channel to comment on the reach of that final state in extracting top polarization. We also comment that, in principle, an improved top-tagging algorithm based on machine learning techniques~\cite{CMS:2019gpd} are expected to improve the results. We plan to take up these issues in follow-up studies.

\section{Acknowledgement}

JL would like to thank Samrat Kadge for his help regarding python programming and useful discussions on the statistical analysis. JL would also like to thank Regional Centre for Accelerator-based Particle Physics (RECAPP), Harish-Chandra Research Institute for cluster facilities.

\bibliographystyle{jhep}
\bibliography{ref1}

\end{document}